\documentclass[12pt,article]{revtex4}
\usepackage{bm}
\usepackage{setspace}
\usepackage[final]{graphicx}
\usepackage{longtable}
\newcommand{\be}{\begin{equation}}
\newcommand{\ee}{\end{equation}}
\newcommand{\bea}{\begin{eqnarray}}
\newcommand{\eea}{\end{eqnarray}}

\newcommand{\eqan}[1]{\begin{eqnarray*}#1\end{eqnarray}}

\begin{document}
\begin{flushleft}
\title{\rmfamily{Investigation of a dilute polymer solution in confined geometries}}
\author{\it{Zoryana Usatenko}}
 \affiliation{Institute of Physics, Faculty of Physics, Mathematics and Computer Science,
 Cracow University of Technology, 30-084 Cracow, Poland}
 \author{\it {Krzysztof S. Danel}}
 \affiliation{Institute of Chemistry, Department of Biopolymer Chemistry,
 University of Agriculture, 30-149 Cracow, Poland}

\begin{abstract}
\begin{flushleft}
\singlespacing
 {\bf Abstract:}
The paper presents a short overview of the theoretical, numerical
and experimental works on the critical behaviour of a dilute polymer
solution of long-flexible polymer chains confined in semi-infinite
space restricted by a surface or in a slit geometry of two parallel
walls with different adsorbing or repelling properties in respect
for polymers as well as in a solution of mesoscopic spherical
colloidal particles of one sort or two different sorts. We discuss
the application of the massive field theory approach in a fixed
space dimensions $d=3$ up to one loop order for such topics as:
$(a)$ the investigation of the elastic properties of real polymer
chain immersed in a good solvent and anchored by one end to the
surface and with other end being under tension of the external
applied force as it is usually takes place in the single - molecule
force spectroscopy experiments; $(b)$ the calculation of the monomer
density profiles, the depletion interaction potentials and the
depletion forces which arise in a dilute polymer solution immersed
in confined geometries, like slit of two parallel walls with
different adsorbing or repelling properties in respect for polymers;
$(c)$ the investigation of monomer density profiles and the
depletion forces which arise in the polymer-colloid mixtures in the
case of the mesoscopic spherical colloidal particles of one sort or
two different sorts. The presented results give possibility better
to understand the complexity of physical effects arising from
confinement and indicate about the interesting and nontrivial
behavior of linear polymer chains in confined geometries and are in
good qualitative and quantitative agreement with previous
theoretical investigations, results of Density Functional Theory
(DFT), Monte Carlo (MC) simulations and experimental data based on
the single molecular atomic force spectroscopy (AFM) and the total
internal reflection microscopy (TIRM). Besides, the obtained results
have important practical applications for better understanding of
the elastic properties of the individual macromolecules, networks,
gels and brush layers as well as indicate about possibility of
application in production of new types of nano- and
micro-electromechanical devices.

\vspace{0.5cm}


\end{flushleft}
\end{abstract}
\maketitle

\begin{figure}[ht!]
\begin{center}
\includegraphics[width=8.0cm]{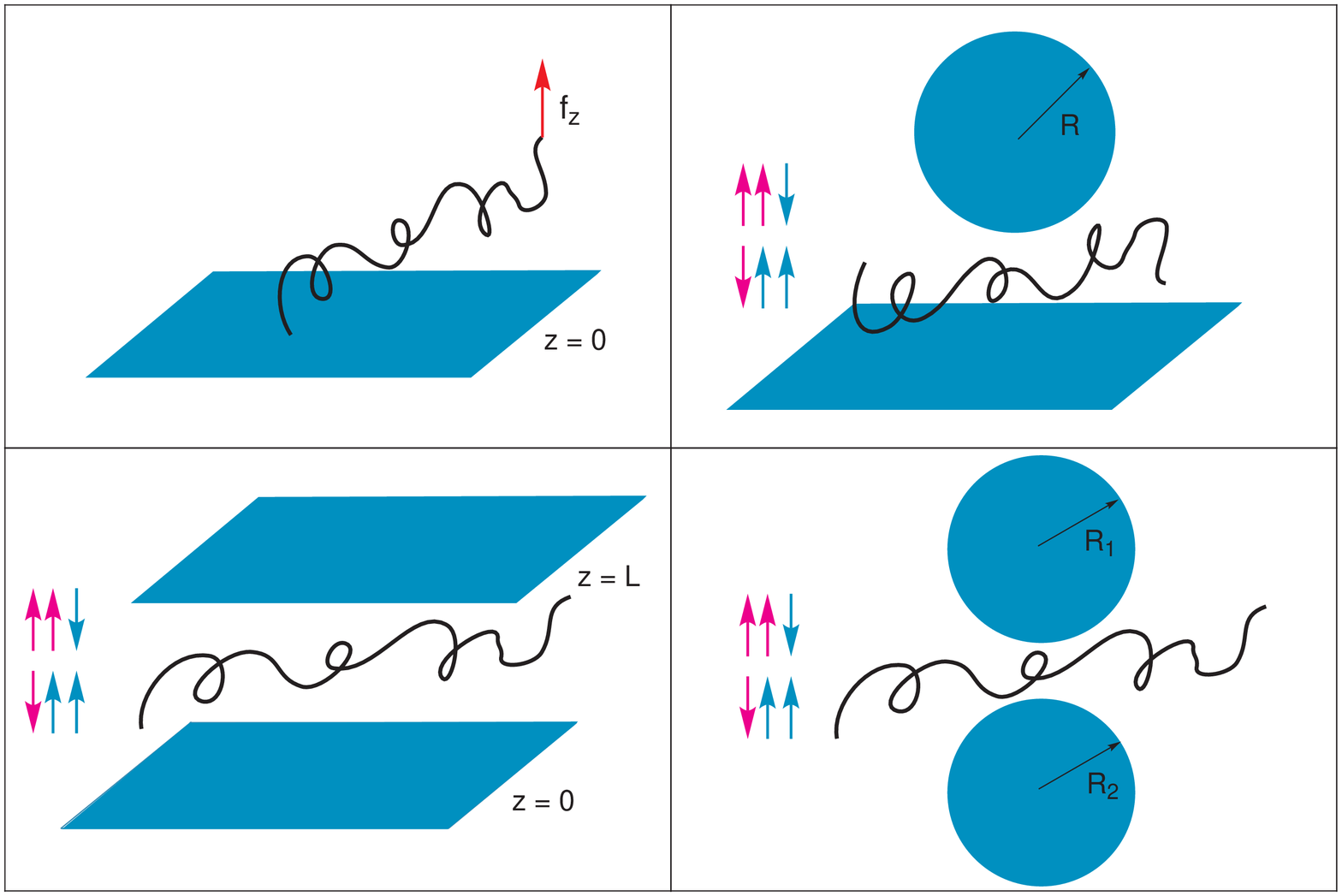}
\label{Fig:1}
\end{center}
\end{figure}

\section*{Introduction}

One of the fundamental questions of the present material science is
the question about adsorption and localization processes of
macromolecular chains on surfaces and in confined geometries as well
as the investigation of the entropic and enthalpic elasticity of
polymer materials and their mechanical stability. This is motivated
by their relevance to numerous practical applications such as
lubrication, adhesion, surface protection, lithography,
chromatography, and biologic phenomena (polymer-membrane
interactions, signal transduction), as well as in biotechnology and
drug delivery. Understanding the basic principles underlying such
processes is of great importance. Besides, the macroscopic
characteristics of polymer materials such as fibers and plastics,
are determined by the mechanical properties of individual polymer
chains. In this respect very important task is the investigation of
the polymer chains deformation due to the influence of external
forces.

As it is known, the single molecular atomic force spectroscopy (AFM)
experiments on DNA, RNA or actin filaments \cite{NLA07,NN08,WR02}
during last years provided information about strains and stresses
which undergo biomolecules during molecular recognition between DNA
and proteins or protein-induced bending of DNA and allowed to
measure the energy transduction during ATP cycle in molecular motors
and to investigate many other fascinating phenomena such as
disentanglement of synthetic polymers
\cite{HS01,HHCMSG02,HHNCROMSG03,UHSGShDXP02,ROHG97} and their
peeling from the surface \cite{HS01,FBJSG04,FSG04,SFJHG03}. As it
was found in \cite{HRSGN05} at low forces the intramolecular
conformations of macromolecules were governed by entropic effects
and at high forces the backbone deformations dominated.

 In the AFM experiments on stretching of
single macromolecules \cite{HRSGN05,KEG06,NHNG06}, one end of a
macromolecule is anchored to the surface and other end is fixed at
the cantilever tip. The tip of cantilever is slowly moved away from
the surface and the corresponding stretching force is measured from
piconewtons up to the femtonewton resolution as a function of the
distance. It should be mentioned that recently a similar procedure
has been applied for measuring the forces induced during the
separation of adsorbed polymers from a surface \cite{HS01,KPG08}.

The development during last years of the experimental techniques,
such as AFM, magnetic beads, optical tweezers or glass microneedles,
which allowed to perform the structural and functional investigation
of macromolecules under tension of external forces stimulated
development of new theoretical approaches and models. In this
contents it should be mentioned very useful the freely jointed chain
(FJC) model \cite{ORG99,HS01,LNK03,RC03}, which was applied for
description of flexible polymer chains deformations and the wormlike
chain (WLC) model \cite{OOPEGM00,KSGB97,HS01,LNK03}, which
frequently was used for description of stiff biopolymers such as a
double - stranded helix of DNA and biological filaments in
unrestricted space. Besides, recently a parameter - free description
of the polymer elasticity based on a combination of statistical
mechanics and quantum mechanics was introduced  in series of papers
\cite{HRSGN05,KWG99} for macromolecules in unrestricted geometry. In
a series of papers \cite{B00,MTM01} devoted to analytical
investigation in the framework of a directed walk model the close
analogy between mechanical desorption of a single polymer chain from
a surface and mechanical unzipping of double-stranded DNA was
mentioned.

 As it is known
\cite{E93,EKB82,WNF87,U06,U11,Ujml11}, the critical behavior of
dilute and semi-dilute  polymer solutions in semi-infinite space
restricted by a surface is more complicated and delicate task which
requires taking into account surface effects and is described by a
new set of surface critical exponents. The series of previous
investigation based on renormalization group (RG) methods
\cite{E93,EKB82,WNF87,U06,UC04,U11,Ujml11} concentrated main
attention on the investigation of the process of polymer adsorption
and calculations of the respective surface critical exponents,
monomer density profiles, number of adsorbed monomers on the surface
and thickness of the adsorbed layer. Besides, the task of
investigation of the stretching and force induced desorption of
anchored to the surface polymer chains which have practical
application in technological process of producing new type of
polymeric materials was the subject of series recent papers
\cite{SKB09,BRSSU13,U14,Umac14}. Such, for example, the theory of
stretching of an end-tethered ideal
 chain (Gaussian chain) anchored to repulsive and inert plane was
proposed in \cite{SKB09}. In one of our previous papers
\cite{BRSSU13} the density functional theory (DFT) was applied for
calculation of the force acting on an end - tethered ideal
(Gaussian) chain and real polymer chain with the excluded volume
interaction (EVI) in a good solvent anchored to repulsive surface.
Unfortunately, the implementation of DFT and Monte Carlo (MC)
techniques are restricted to the case of rather short polymer chains
due to the computational difficulties.

It stimulated us to apply the massive field theory approach (MFT)
for investigation of stretching of long - flexible real polymer
chain with the EVI in a good solvent anchored by one end to
repulsive or inert surface \cite{U14,Umac14}. It should be mentioned
that initially MFT approach was proposed by Parisi for infinite
systems in \cite{Par80} and subsequently applied to semi-infinite
systems by Diehl and Shpot in \cite{DSh98}. Recently
 we applied the massive field theory approach in a fixed space dimension $d=3$
 for investigation of a dilute polymer solution of real polymer chains with the EVI in a good
solvent confined in a slit geometry of two parallel walls with
different boundary conditions \cite{RU09,RUIOP09} by analogy as it
was proposed in \cite{ShHKD01} for two repulsive walls in the
framework of the dimensionally regularized continuum version of the
field theory with minimal subtraction of poles in $\epsilon=4-d$ -
expansion, where d - is dimensionality of space. Such, for example,
in \cite{RU09,RUIOP09} we investigated the situation of two
repulsive walls (Dirichlet-Dirichlet boundary conditions (b.c.)) two
inert walls (Neumann-Neumann b.c.) and mixed case of one repulsive
and the other one inert wall (Dirichlet-Neumann b.c.). Taking into
account the Derjaguin approximation \cite{D34} we obtained the
results for the depletion interaction potentials and the depletion
forces for the case of one mesoscopic colloidal particle of big size
near the wall and for the case of two big colloidal particles with
different adsorbing or repelling properties in respect to polymers
\cite{RU09,RUIOP09,UKChR17} and compared the obtained analytical
results with experimental results obtained by Rudhardt, Bechinger
and Leiderer \cite{RBL98}.

Besides, in a series of our papers \cite{U11,Ujml11} the universal
density-force relation was analyzed by analogy as it was proposed by
Joanny, Leibler, de Gennes \cite{JLG} and by
 Eisenriegler \cite{E97}. Such, for example, in \cite{U11,Ujml11} the corresponding universal
amplitude ratio and the monomer density profiles of linear ideal and
real polymer chains with the EVI in a good solvent immersed in a
slit geometry of two parallel repulsive walls, one repulsive and the
other one inert wall were obtained in the framework of the massive
field theory approach directly in fixed space dimensions $d=3$.
Taking into account the Derjaguin approximation \cite{D34}, the
monomer density profiles of a dilute polymer solution confined in a
semi-infinite space containing the mesoscopic spherical colloidal
particle of big size with different adsorbing or repelling
properties in respect for polymers were calculated
\cite{U11,Ujml11}. It should be mentioned, that the interaction of
long flexible nonadsorbing linear polymers with mesoscopic colloidal
particles of big and small size and different shape was the subject
a series of papers \cite{BEShH99,HED99} based on $\epsilon=4-d$ -
expansion. The obtained in a series of papers
\cite{BEShH99,HED99,U11,Ujml11,RU09,UKChR17} results for long
flexible linear polymer chains indicate that focusing on such
systems leads to universal results which are independent of
microscopic details and are free of nonuniversal model parameters
and depend only on shapes of particles, adsorbing or repelling
properties of particles in respect to polymers and ratios of three
characteristic lengths of the system such as the radius of the
particle, the polymer size, the distance between the particle and
the wall or between two particles, respectively.

The calculation of the depletion interaction potential and the
depletion force between two parallel walls in a slit geometry, or
between two colloidal particles (or wall and a single colloidal
particle) of arbitrary shape was performed also by other theoretical
methods including the method of hard spheres\cite{AO54,AO58} and the
self-consistent field theory \cite{O96,O97}. It should be mentioned
that the theoretical investigations were verified by numerical
methods \cite{MB98,HG04} using the model of RW for ideal polymer
chain at $\Theta$ - solvent and SAW for a real polymer chain with
the EVI in a good solvent confined inside the slit of two repulsive
walls.

In order to give answer for the question why we use for calculations
the massive field theory approach \cite{Par80,Z89,SD89,DSh98} we
should notice that the above mentioned approach works directly in a
fixed space dimensions $d=3$ and avoids the $\epsilon=4-d$ -
expansion. In this respect we should also recollect that in the
dimensionally regularised continuum version of the field theory
based on the $\epsilon=4-d$ - expansion for fixed $d<4$ space
dimensions arise infrared singularities, which manifest themselves
as poles in Feynman integrals at rational values of $\epsilon$  and
accumulate at the upper critical dimension $d^{*}=4$ as the order of
perturbation theory increases \cite{S73,BD82,DD81,E93,EKB82,WNF87}.
In accordance with it arises the problem of summing up these
infrared singularities, which arise at small momenta and long
wavelength. As was mentioned by Parisi \cite{Par80} it should be
introduced some additional hypothesis on the summation of these
singularities, because any calculations based on the $\epsilon$ -
expansion and the RG in this perturbative zero-mass scheme do not
contain any information about the critical behavior in the case of
fixed dimension $d<d^{*}$. One of such approaches, which works
perfectly directly in a fixed space dimension and avoid the
$\epsilon$ - expansion is the massive field-theory approach for
fixed space dimension $d<d^{*}$ \cite{Z89,Par80,SD89,DSh98} which we
used for calculations in our papers
\cite{U06,UC04,RU09,RUIOP09,U11,Ujml11,UH17,UHK16,UKChH17,UKChR17}.

\section*{The model}

As it was mentioned in a series of papers
\cite{deGennes,deGennes79,CJ90,E93}, the behaviour of long flexible
polymer chains is determinated by entropy and characterized by
detail independence, universality and scaling. The properties of
scaling and universality also take place for critical systems with
surfaces \cite{B83,D86} and can be applied for the case of polymer
solution in confined geometries.

In a series of our papers (see
\cite{U06,RU09,RUIOP09,U11,Ujml11,U14,Umac14,UKChR17}) the
investigations of a dilute solution of linear polymer chains
confined in semi-infinite space restricted by a surface or in a slit
geometry of two parallel walls and in a solution of mesoscopic
spherical colloidal particles of one sort or two different sorts
were performed. We assumed that the solution of polymers is
sufficiently dilute so that the interchain interactions and the
overlapping between different polymers can be neglected and it is
sufficient to consider the behaviour of a single polymer chain.

 As it is known, the behavior of a single ideal (Gaussian) polymer chain at $\theta$-solvent can be
described by a model of random walk (RW) and the behavior of
long-flexible real polymer chain with the EVI in a good solvent for
temperatures above the $\theta$-point by a model of self-avoiding
walk (SAW). Usually in the case when the EVI between monomers
becomes relevant the polymer coils are less compact than in the case
of ideal chains. Taking into account the polymer-magnet analogy
developed by de Gennes \cite{deGennes,deGennes79}, we can switch
from an Edwards type model for long - flexible real polymer chains
with the EVI in a good solvent to the field theoretical $\phi^4$
$O(n)$- vector model in the formal $n \to 0$ limit at the infinite
number of steps (monomers) $N$. The value $1/N$ plays the role of a
critical parameter analogous to the reduced critical temperature in
magnetic systems. In the case when the polymer solution is in
contact with
 solid substrates, the interaction between the monomers and the surfaces should be taken into account.
As it was noted by de Gennes \cite{deGennes76} and by Barber et al.
\cite{Barber}, there is a formal analogy between the polymer
adsorption problem and the equivalent problem of critical phenomena
in the semi-infinite $|\phi|^4$ $O(n)$-vector model of a magnet with
a free surface \cite{DD81,D86}.
  In a series of papers \cite{BJ05,MBJ09,MJB10} the extensive multicanonical
 MC computer simulations were used in order to study the conformational
 behaviour of flexible polymers near an attractive
 surfaces and the corresponding complete solubility-temperature phase diagram was
 constructed and discussed.

The proposed in
\cite{E93,ShHKD01,U06,RU09,RUIOP09,U11,Ujml11,U14,Umac14,UKChR17}
calculations were performed for the case when the surfaces were
impenetrable. It means that the corresponding potential $U(z)$ of
the interaction between the monomers of a polymer chain and a wall
tends to infinity $U(z)\to\infty$ when the distance between a wall
and polymer chain is less than monomer size $l$. The deviation from
the adsorption threshold $(c\propto(T-T_a)/T_a)$ (where $T_{a}$ is
adsorption temperature) changes sign at the transition between the
adsorbed (the so-called normal transition, $c<0$) and the
nonadsorbed state (ordinary transition, $c>0$) \cite{D86,E93} and it
plays the role of a second critical parameter. The value $c$
corresponds to the adsorption energy divided by $k_{B}T$ or the
surface enhancement in field theoretical treatment. It should be
mentioned that the adsorption threshold for long-flexible polymer
chains takes place when $1/N\to 0$ and $c\to 0$. In the case when a
polymer solution is confined in a slit geometry of two parallel
walls the properties of the system depend on the ratio $L/\xi_R$,
where $L$ is the distance between two walls and
$\xi_R=\sqrt{<R^2>}\sim N^{\nu}$ is the average
 end-to-end distance, as it was shown in a series of papers \cite{ShHKD01,RU09,RUIOP09,U11,Ujml11}.

 Taking into account the polymer-magnet analogy \cite{deGennes,deGennes79}, the
respective partition function $Z({\bf{x}},{\bf{x'}})$ of a single
polymer chain with two ends fixed at ${\bf{x}}$ and ${\bf{x'}}$ is
connected with the two-point correlation function
$G^{(2)}({\bf{x}},{\bf{x'}})=<{\vec{\phi}({\bf x})}{\vec{\phi}}({\bf
x}')>$ in  $\phi^4$ $O(n)$- vector model for $n$-vector field
$\vec{\phi}({\bf{x}})$ with the components $\phi_i(x)$, $i=1,...,n$
(and ${\bf{x}}=({\bf r},z)$) via the inverse Laplace transform
$\mu_{0}^{2}\to L_{0}$: $Z({\bf x},{\bf x}';N,v_{0})={\cal
IL}_{\mu_{0}^2\to L_{0}}(<{\vec \phi}({\bf x}){\vec\phi}({\bf
x'})>|_{n\to 0})$ in the limit, where the number $n$ of components
tends to zero \cite{deGennes,deGennes79,CJ90,E93,RU09,U11}. The
conjugate Laplace variable $L_{0}$ has the dimension of length
squared and is proportional to the total number of monomers $N$ of
the polymer chain: $L_{0}\sim {l^2}N$.

The effective Ginzburg-Landau-Wilson Hamiltonian describing the
system of a dilute polymer solution confined in semi-infinite
($i=1$) or confined geometry of two parallel walls ($i=1,2$) is
\cite{D86}: \be {\cal H}[{\vec \phi,\mu_{0}}] =\int
d^{d}x\bigg\lbrace \frac{1}{2} \left( \nabla{\vec{\phi}} \right)^{2}
+\frac{{\mu_{0}}^{2}}{2} {\vec{\phi}}^{2} +\frac{v_{0}}{4!}
\left({\vec{\phi}}^2 \right)^{2} \bigg\rbrace
+\sum_{i=1}^{2}\frac{c_{i_{0}}}{2} \int d^{d-1}r {\vec{\phi}}^{2},
\label{hamiltonianslit}\ee where the variables $\mu_{0}$ plays the
role of chemical potentials (or "bare mass" in field-theoretical
treatment), $v_0$ is the "bare coupling constant" which
characterizes the strength of the EVI between any monomers of
polymer chain. The case $v_{0}=0$ corresponds to the situation of an
ideal polymer chain when only monomer - monomer interactions between
consecutive monomers along the chain take place. In the case of a
slit geometry the walls are located at the distance $L$ one from
another in $z$- direction in such way that the surface of the bottom
wall is located at $z=0$ and the surface of the upper wall is
located at $z=L$. Each of the two system surfaces is characterized
by a certain surface enhancement constant $c_{i_{0}}$, where
$i=1,2$. The model defined in (\ref{hamiltonianslit}) is restricted
to translations parallel to the boundaring surfaces. Thus, only
parallel Fourier transformations in $d-1$ dimensions with respect to
"parallel" coordinates ${\bf{r}}$ take place.

The interaction between the polymer chain and the walls is
implemented by the different boundary conditions. In a series of our
papers we performed investigations for different boundary
conditions. Such, for example, when we discussed the processes of
adsorption and desorption on the surface of a dilute solution of
polymer chains \cite{U06} and stretching anchored to the surface
long-flexible real polymer chain with the EVI in a good solvent for
temperatures above the $\theta$-point \cite{U14,Umac14}, we
performed calculations for the case of repulsive wall and wall where
the adsorption threshold takes place (so-called \cite{E93} "inert"
wall). In the case of repulsive wall (i.e. where the segment
partition function and thus the partition function for the whole
polymer chain $Z({\bf x},{\bf x}';L_{0})$ tends to 0 as any segment
approaches the surface of the wall: $z, z'\to 0$) Dirichlet boundary
condition takes place: $ {\vec \phi}({\bf{r}},z)=0$ at $z=0$. This
corresponds to the fixed point boundary condition: $c\to +\infty$ of
the ordinary transition of the field theory. In the case of inert
wall the fields $\vec{\phi}({\bf{r}},z)$ satisfy Neumann boundary
conditions: $ \frac{\partial{\vec \phi}({\bf {r}},z)}{\partial
z}|_{z=0}=0$. The requirement, describing the inert wall corresponds
to the fixed point $c=0$ of the so-called special transition
\cite{DD81,D86,DSh98} in the field theoretical treatment.

Besides, in a series of papers
\cite{RU09,RUIOP09,U11,Ujml11,UKChR17} we performed investigation of
a dilute polymer solution of linear polymer chains immersed in a
slit geometry of two repulsive walls, two inert walls, and for the
mixed case of one repulsive and the other one inert wall. The
situation of two repulsive walls corresponds to the
Dirichlet-Dirichlet boundary conditions (D-D b.c.)( see also
\cite{ShHKD01,RU09,RUIOP09}): \be {\vec \phi}({\bf {r}},0)={\vec
\phi}({\bf {r}},L)=0\quad or\quad c_{1}\to +{\infty},\quad c_{2}\to
+{\infty}\label{DD}. \ee The case of two inert walls
  corresponds to the Neumann-Neumann boundary
 conditions (N-N b.c.) (see also \cite{RU09,RUIOP09}):
\be \frac{\partial{\vec \phi}({\bf {r}},z)}{\partial
z}|_{z=0}=\frac{\partial{{\vec \phi}({\bf {r}},z)}}{\partial
z}|_{z=L}=0\quad or\quad c_{1}=0,\quad c_{2}=0 \label{NN},\ee and
for the mixed case of one repulsive and the other one inert wall,
the Dirichlet-Neumann boundary conditions(D-N b.c.) takes place
\cite{RU09,RUIOP09}: \be {\vec \phi}({\bf {r}},0)=0,
\quad\frac{\partial{{\vec \phi}({\bf {r}},z)}}{\partial
z}|_{z=L}=0\quad or \quad c_{1}\to +{\infty},\quad
c_{2}=0\label{DN}.\ee  The performed calculations are valid for the
case of wide slit limit $y\geq 1$ (where $y=\frac{L}{R_{x}}$), but
they are not suitable to describe the case of dimensional crossover
from $d$ to $d-1$ dimensional system which arises for $y<<1$. The
$d-1$ dimensional system is characterized by another critical
temperature and a new critical fixed point, as it takes place, for
example, in magnetic or liquid thin films. Nevertheless, some
assumptions allowed us to describe the region of narrow slit, as it
was proposed in one of our papers (see Ref.\cite{RU09}). In the case
of infinitely large wall separations, the slit system decomposes
into two half-space systems.

It should be mentioned, that the only relevant length of the model
is the average end-to-end distance $<{\bf{R}}^{2}>\sim N^{2\nu} $,
and
$<R_{x}^{2}>_{bulk}=<R_{y}^{2}>_{bulk}=<R_{z}^{2}>_{bulk}=<\frac{{\bf{R}}^{2}}{d}>_{bulk}
$ where $ R_{x},R_{y},R_{z} $ are the projection of the end to end
distance $ {\bf R} $ onto the direction of $x,y,z$ axis and $\nu$ -
is Flory exponent which is $1/2$ for ideal polymer chain and 0.588
for real polymer chain with the EVI.

\section*{Calculation of the average stretching force}
A series of papers \cite{SKB09,BRSSU13,U14,Umac14} were devoted to
the investigation of stretching anchored to the surface
long-flexible linear polymer chain with the EVI in a good solvent
for temperatures above the $\theta$-point. The difference in the
strain-force and the force-strain dependences, which corresponds to
the $f$ ensemble and $Z$ ensemble, respectively, was discussed in
\cite{SKB09}. As it was mentioned in \cite{SKB09}, in $Z$ ensemble
stretching usually was performed via an increase in the end-to-end
distance and in the case of $f$ ensemble the stretching was provided
by the external force field. It was established, that the relation
between strain and force is similar for the both ensembles for the
case of ideal polymer chain and for the case of real polymer chain
with the EVI in a good solvent the difference diminishes in the
limit of infinite number of monomers $N\to\infty$.

 In general, the average stretching force which
is applied to the free end (which is situated in the layer on the
distance $z'$ from the surface) of the
 polymer chain anchored by other end to the surface at $z=0$ can be calculated
 and has a form: \be
\frac{f(z')}{k_{B}T}=-\frac{\partial}{\partial z'}ln Z (z=0,z'; N).
\label{Pincus-force}\ee
 This force is analogous to the well known Pincus
force \cite{deGennes79}.

 Such in the case of ideal
polymer chain anchored by one end to repulsive surface the result
for the average stretching force has a form \cite{U14,Umac14}: \be
\frac{<f^{id}_{ord}({\zeta})>R_{z}}{k_{B}T}=-\frac{1}{\zeta}(1-{{\zeta}}^{2}),\label{Pincus-ideal-ord}
\ee where ${\zeta}=\frac{z'}{R_{z}}$. The case of real polymer
chains is more complicated, because the EVI should be taken into
account.  Thus,
 after performing the mass $\mu_{0}^{2}=\mu^{2}+\delta \mu$
 and the surface enhancement
 $c_{0}=c+\delta c$ renormalizations of the correspondent correlation
 functions and after vertex normalization of the coupling constant $v_{0}=v\mu$,
 as it is usually accepted in the massive field-theory approach of semi-infinite systems in
fixed space dimensions $d=3$ (see
 \cite{DSh98,UShH01,U11}), the respective result for the average stretching force applied to
free end of real polymer chain with the EVI in a good solvent
anchored by other end to repulsive surface was obtained
\cite{U14,Umac14} in accordance with  Eq.(\ref{Pincus-force}): \be
\frac{<f^{real}_{ord}({\zeta})>R_{z}}{k_{B}T}=-\frac{1}{\zeta}(1-{\zeta}^2+\frac{\tilde{v}^{*}}{4}e^{-4{\zeta}^{2}}
(1+\frac{1}{4{\zeta}^{2}})).\label{Pincus-real-ord}\ee It should be
mentioned that the calculations were performed at the correspondent
stable fixed point ${\tilde{v}}^{*}=1$ obtained from resummed beta
functions $\beta({\tilde{v}})$ of the underlying bulk field theory
in one-loop order.

In a similar way were performed the calculations for the average
stretching force applied to free end of polymer chain anchored to
inert surface. Thus, for the case of ideal chain we obtained
\cite{U14,Umac14}: \be
\frac{<f^{id}_{sp}(\zeta)>R_{z}}{k_{B}T}={\zeta}.\label{Pincus-ideal-sp}
 \ee
The obtained in Eq.(\ref{Pincus-ideal-sp}) result indicates that in
the case when free end of polymer chain moves away from the inert
surface, the average stretching force increases linearly. Taking
into account the massive field - theory approach at fixed $d=3$
dimensions \cite{DSh98,UH02}, the correspondent average stretching
force applied to free end of real polymer chain with the EVI in a
good solvent anchored by other end to inert surface was obtained in
\cite{U14,Umac14}:
 \be
\frac{<f^{real}_{sp}({\zeta})>R_{z}}{k_{B}T}={\zeta}(1+\frac{\tilde{v}^{*}}{16}).\label{Pincus-real-sp}
\ee

  It should be mentioned that the results obtained
in the framework of the present calculation scheme for the average
stretching force in the case of ideal polymer chain anchored to
repulsive and inert surfaces (see Eq.(\ref{Pincus-ideal-ord}) and
Eq.(\ref{Pincus-ideal-sp})) coincide with the results obtained by
Skvortsov et. al \cite{SKB09} for Gaussian chain model. The
comparison of our results Eq.(\ref{Pincus-ideal-ord}) and
Eq.(\ref{Pincus-ideal-sp}) (see also \cite{U14,Umac14}), and the
results obtained by Skvortsov et. al \cite{SKB09} is possible to
observe in Figure 1. Our results for the stretching force applied to
anchored to the repulsive surface ideal chain marked as black and
blue dashed lines for number of monomers N=50 and N=500,
respectively, and the results obtained by Skvortsov et al. in
\cite{SKB09} for Gaussian chain are shown as orange triangles and
green circles, respectively. We took into account that the
projection of the end-to-end distance on the $z$ axis is:
$R_{z}^{2}=<R_{z}^{2}>_{bulk}=\frac{b^{2}N^{2\nu}}{3}$ and $\nu=1/2$
for ideal chain, $b$ is the effective segment length of the polymer
model under consideration. It should be mentioned, that Figure 1
contains also our results obtained in the framework of the massive
field theory (MFT) approach at fixed space dimensions d=3 for the
case of real polymer chain with the EVI in a good solvent with N=50
(orange solid line) and with N=500 (green solid line) anchored to
repulsive surface. Unfortunately, in \cite{SKB09} the situation of
real polymer chains anchored to the surface was not discussed and
only the task of finding stretching force applied to the ends of
real polymer chain in a good solvent immersed in unrestricted space
was discussed.
\begin{figure}[ht!]
\begin{center}
\includegraphics[width=8.0cm]{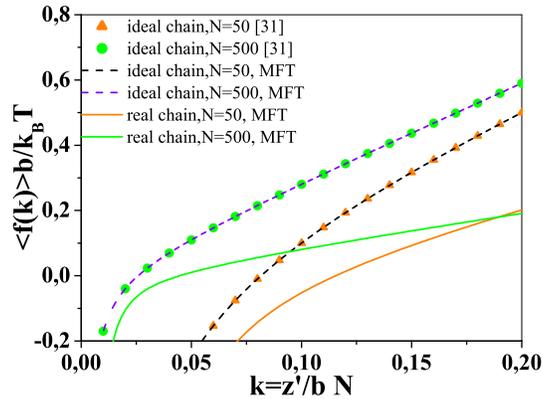}
\caption{Comparison of the massive field theory (MFT) results
obtained for the average stretching force applied to free end of
ideal and real polymer chain anchored to repulsive surface as
function of $k=z'/b N$ and the results obtained by Skvortsov et
al.\cite{SKB09} for Gaussian chain for number of monomers $N=50$ and
$N=500$. The results obtained in the framework of both methods for
ideal chain (Gaussian chain) coincide.} \label{Fig:1}
\end{center}
\end{figure}
In a similar way we compared our results with the ones obtained in
\cite{SKB09} for the average stretching force applied to an ideal
chain anchored to inert surface and found complete agreement (see
\cite{U14,Umac14}).
 As it is known \cite{deGennes79,RC03,SKB09}, the force extension relation
for real polymer chain with the EVI in a good solvent without
surface reduces to the Pincus deformation law: $z\sim N(\frac{f
l}{k_{B}T})^{\frac{1-\nu}{\nu}}$. The presence of the surface should
modify the force extension relation. For example, from the results
obtained in the framework of the dimensionally regularised field
theory with $\epsilon$-expansion in \cite{WNF87}  for the case of
real polymer chain immersed in a slit of two parallel walls with
different boundary conditions and anchored by one or two ends to
surface follows, that the correspondent stretching force will be
proportional to $1/z'$ for both cases of Neumann and Dirichlet
boundary conditions. This result is in disagreement with the result
obtained by \cite{SKB09} for the case of ideal (Gaussian) chain
anchored to inert surface (Neumann b.c.), where average force is
proportional to extension $z'$. In \cite{E97} was mentioned, that
the force extension relation for the case of polymer chain anchored
to the surface should be different for two limiting cases
$z'<<R_{z}$ and $z'>>R_{z}$. But, unfortunately in \cite{E97}
detailed analysis for these cases was not performed. The detailed
analysis of our results, obtained in Eq.(\ref{Pincus-real-ord})
shows, that in the case $z'<<R_{z}$ (or ${\zeta}<< 1$) we obtain
that the average stretching force $\frac{<f({\zeta})>}{k_{B} T}$ for
long - flexible real polymer chain anchored to the repulsive surface
is proportional to $1/z'$, which is in agreement with scaling
arguments of Flory theory and theory of Pincus blobs adsorbed on the
surface \cite{F53,RC03}. In the intermediate case, when $z'=R_{z}$
the $1/z'$ dependence of the stretching force is still valid. In the
case $z'>>R_{z}$ (or ${\zeta}>>1$) we come to the dependence
$\frac{<f({\zeta})> R_{z}}{k_{B}T}\sim {\zeta}^{\frac{\nu}{1-\nu}}$.
The analysis of our results for the average stretching force in the
case of real polymer chain anchored to inert surface
Eq.(\ref{Pincus-real-sp}) is in agreement with the result obtained
in \cite{SKB09} for Gaussian chain anchored to inert surface, where
the average stretching force is proportional to extension $z'$.
Besides, we observed good qualitative agreement of the obtained
analytical results with experimental results based on the AFM
measurements \cite{HS01,NHNG06} as it was shown in \cite{U14}.

In \cite{Umac14} the comparison of the analytical results obtained
in the framework of the massive field theory approach and the
results obtained in the framework of DFT approach and Monte Carlo
(MC) calculations \cite{BRSSU13} were performed (see Figure 2$(a)$
and Figure 2$(b)$). It should be mentioned that in order to do this
comparison we introduced some changes of variables, as it was
mentioned in \cite{Umac14}. We assumed, that $t=z'/N$, then the
result obtained in Eq.(\ref{Pincus-ideal-ord}) for the average
stretching force has the form: \be
\frac{<f^{ideal}_{ord}(t)>}{k_{B}T}=-\frac{1}{N t}(1-\frac{t^{2} N
d}{b^{2}}).\label{comp1} \ee  The correspondent expression for the
average stretching force in the case of real polymer chain with the
EVI anchored by one end to the repulsive surface as function of
$t=z'/N$ is: \be \frac{<f^{real}_{ord}(t)>}{k_{B}T}=-\frac{1}{N
t}(1-\frac{t^{2} A d}{b^2}+\frac{\tilde{v}^{*}}{4}e^{-\frac{4 t^{2}A
d}{b^2} }(1+\frac{b^2}{4 t^{2} A d})),\label{comp2} \ee where
$A=\frac{N^2}{N^{2\nu}}$ and $A\approx6.67$ for N=10,
$A\approx11.81$ for N=20. Figures 2$(a)$ and 2$(b)$ present
comparison of the obtained in the framework of the massive field -
theory approach results for the average stretching force (see
Eq.(\ref{comp1}), Eq.((\ref{comp1}))) and the results obtained in
the framework of DFT approach and MC calculations \cite{BRSSU13} for
different numbers of monomers N=10 and N=20, respectively.
\begin{figure}[ht!]
\begin{center}
\includegraphics[width=7.0cm]{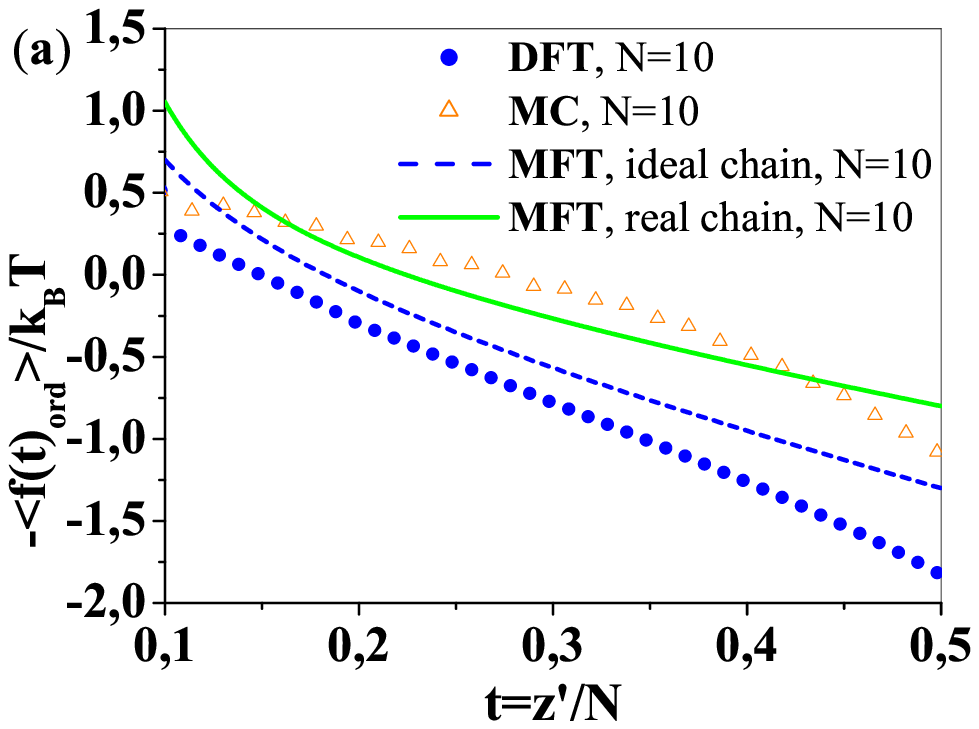}
\includegraphics[width=7.2cm]{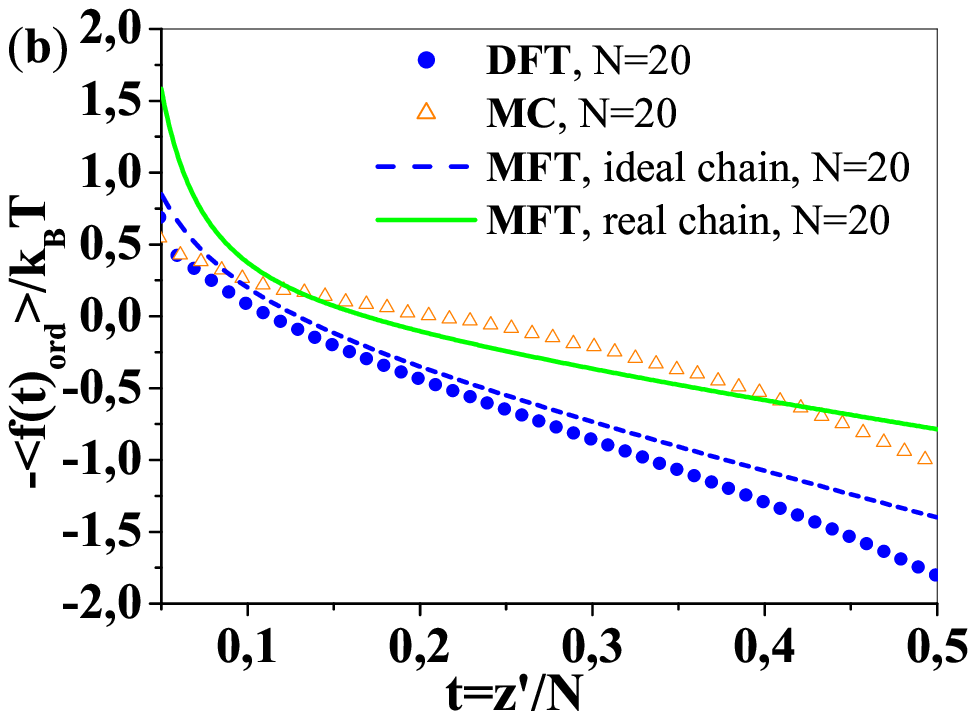}
\caption{Comparison of the massive field theory results obtained for
the average stretching force applied to free end of ideal and real
polymer chain anchored to repulsive surface as function of $t=z'/N$
and the results of DFT  and MC simulations \cite{BRSSU13} for: $(a)$
$N=10$; $(b)$ $N=20$ with $b$=1} \label{Fig:2}
\end{center}
\end{figure}
The comparison of the results presented in Figures 2$(a)$ and 2$(b)$
indicates that the obtained in the framework of DFT and MC methods
results are in good agreement with the analytical results obtained
with use the massive field theory approach for the region of small
applied forces when deformation of polymer chain is not bigger than
linear extension of polymer chain in this direction, $z'/R_{z}\leq
1$. Besides, the better agreement between our analytical results and
the results obtained in the framework of DFT and MC is observed in
the case, when the number of monomers increases and polymer chain
becomes longer. It should be mentioned, that in the case of strong
deformations of polymer chain the Gaussian distribution does not
give good results for the stretching force and in this case the
calculations should be performed with taking into account Langevin
distribution function. This task is more complicated and should be
investigated separately, such as requires more delicate treatment,
specially for the case of real polymer chains anchored to the
surface. In general, it should be mentioned, that the obtained
results have important practical applications for better
understanding of the elastic properties of the individual
macromolecules, networks, gels and brush layers.

\section*{Dilute solution of ideal and real polymer chains in a slit geometry of two parallel walls}

The task of investigation of a dilute solution of ideal and real
polymer chains with the EVI in a good solvent immersed in a slit
geometry of two parallel walls situated at a distance $L$ one from
another with different boundary conditions was the subject of series
papers \cite{ShHKD01,RU09,RUIOP09}. The thermodynamic description of
the problem proposed in \cite{ShHKD01} assumed that the polymer
solution in the slit is in equilibrium contact with an equivalent
solution in the reservoir outside the slit and allowed for the
exchange of polymer coils between the slit and the reservoir. As it
was shown in \cite{ShHKD01,RU09}, the free energy of the interaction
between the walls in such a grand canonical ensemble is defined as
the difference of the free energy of an ensemble where the
separation of the walls is fixed at a finite distance $L$ and where
the walls are separated infinitely far from each other: \be \delta F
= -k_B T \,{{\tilde{N}}}\,\ln\left(\frac{{\cal Z}_{||} (L)}{{\cal
Z}_{\parallel}(L\to\infty)}\right)\,=\, -k_B T \,{
\tilde{N}}\,\bigg\lbrace\ln\left(\frac{{\cal Z}_{||}(L)}{{\cal
Z}}\right) -\ln\left(\frac{{\cal Z}_{||}(L\to\infty)}{{\cal
Z}}\right)\bigg\rbrace, \label{Fgca}\ee where ${\tilde{N}}$ is the
total amount of polymer coils in the solution and $T$ is the
temperature. The ${\cal Z}_{\parallel}(L)$ value is the partition
function of one polymer located in a volume $V$ containing two walls
at the distance $L$. It should be mentioned that the partition
functions ${\cal Z}_{\parallel}(L)$ and ${\cal
Z}_{\parallel}(L\to\infty)$ were normalized on the function ${\cal
Z}=V{\hat{\cal{Z}}}_{b}$, where
${\hat{\cal{Z}}}_{b}={\cal{IL}}_{\mu_{0}^{2}->\frac{R_{x}^2}{2}}[\frac{1}{\mu^{2}_{0}}].$

As it was mentioned in \cite{ShHKD01,RU09}, the corresponding
reduced free energy of interaction $\delta f$ per unit area $A=1$
for the case of linear polymers confined in the slit geometry of two
parallel walls after performing Fourier transform in the direction
parallel to the surfaces and integration over $d^{d-1}r$ can be
obtained:
 \be \delta f =  \frac{\delta{F}}{n_{p}k_{B}T}. \label{dfringz} \ee
 Here $n_p={{\tilde{N}}}/V$ is the polymer
density in the bulk solution. The reduced free energy of interaction
$\delta f$, according to Eq.(\ref{dfringz}), is a function of the
dimension of length and dividing it by another relevant length
scale, for example, the size of polymer in bulk, e.g. $R_x$ yields a
universal dimensionless scaling function for the {\it depletion
interaction potential}:
 \be \Theta(y) = ~\frac{\delta
f}{R_x},\label{Theta}\ee where $ y = L/R_x $ is a dimensionless
scaling variable. The resulting scaling function for the depletion
force between two walls induced by the polymer solution is denoted
as: \be\Gamma(y) =~-\frac{d (\delta
f)}{dL}~=~-\frac{d\Theta(y)}{dy}. \label{Gamma}\ee

In \cite{RU09,RUIOP09} during the investigation of ideal and real
polymer chains with the EVI in a good solvent immersed in a slit
geometry of two parallel walls with D-D b.c. Eq.(\ref{DD}) (the case
of two repulsive walls) we obtained that if both $c_1$ and $c_{2}$
are positive, the depletion interaction potential ${\Theta}_{DD}(y)$
is negative and hence the walls attract each other due to the
depletion zones near repulsive walls. It should be mentioned, that
the obtained in the framework of the massive field theory approach
at fixed space dimensions $d=3$ results \cite{RU09,RUIOP09} for the
dimensionless scaling function of the depletion interaction
potential and the depletion force for ideal and real polymer chains
in a slit geometry of two repulsive walls are in good quantitative
agreement with the previous theoretical results obtained in
Ref.\cite{ShHKD01} via using the dimensionally regularized continuum
version of the field theory with minimal subtraction of poles in
$\epsilon=4-d$ expansion (see Figure 3$(a)$). The obtained
analytical results indicate that the reduction in the depletion
effect due to the EVI is weaker within the massive field theory
approach as compared to the results obtained using the $\epsilon$ -
expansion. For the dimensionless scaling function of the depletion
interaction potential in the case of ideal polymer chain immersed in
a slit geometry of two inert walls, which corresponds to N-N b.c.
Eq.(\ref{NN}) in \cite{RU09,RUIOP09} we obtained
${\Theta}_{NN}(y)=0$. It corresponds to the fact that ideal chains
do not loose free energy inside the slit in comparison with free
chains in unrestricted space. The entropy loss is fully regained by
surface interactions provided by two walls. In the case of real
linear polymer chain with the EVI in a good solvent immersed in a
slit geometry of two inert walls we obtained (see
\cite{RU09,RUIOP09}) that resulting force becomes repulsive due to
excluded volume effects between monomers of polymer chain. In the
case of one repulsive and the other one inert wall, which
corresponds to D-N b.c. Eq.(\ref{DN}) we obtained in \cite{RU09}
that $ \Theta_{DD}(2y){\approx} 2 \Theta_{DN}(y)$ for ideal polymer
chains. This result is intuitively clear, because the depletion zone
is formed only near the lower wall, i.e. near the wall with
Dirichlet b.c. The upper wall with Neumann b.c. does not contribute
at all to the induced depletion interaction. Thus, in
\cite{RU09,RUIOP09} the results for the dimensionless scaling
functions of the depletion interaction potential and the depletion
force were obtained  for ideal and real linear polymer chains with
the EVI in a good solvent confined in a slit geometry of two
parallel walls with D-N b.c. Besides, in \cite{RU09,RUIOP09} the
asymptotic region of wide slit with $y\geq 1$ and narrow slit with
$y<<1$ were discussed.

\begin{figure}[h]
\begin{center}
\includegraphics[width=7.0cm]{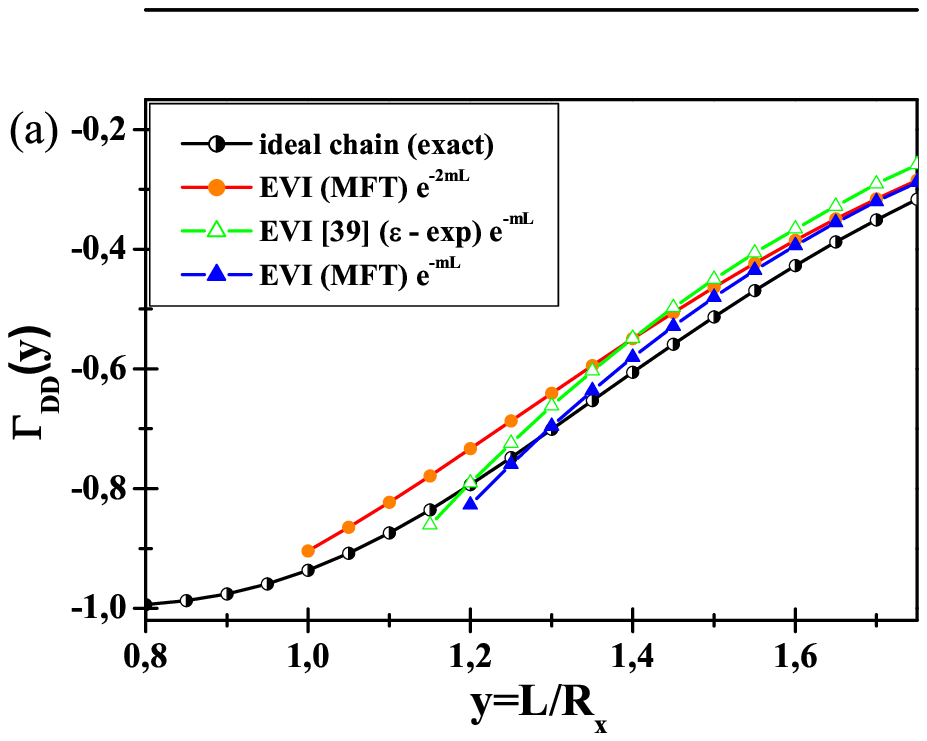}\hspace{0.1cm}
\includegraphics[width=7.0cm]{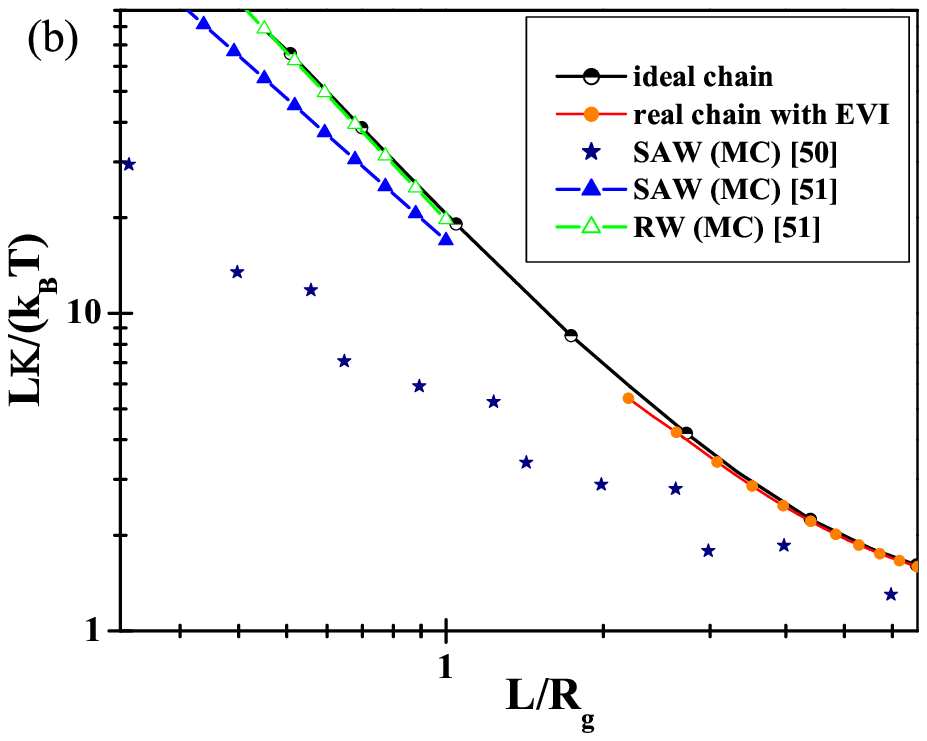}\hspace{0.1cm}
\caption{$(a)$ The dimensionless scaling functions of
$\Gamma_{DD}(y)$ for linear polymer chain in a slit geometry of two
repulsive walls (see\cite{RU09,RUIOP09}) in comparison to the
results obtained in \cite{ShHKD01}; $(b)$ Comparison of the
analytical results obtained in \cite{RU09,RUIOP09} for the reduced
canonical force and the results obtained by MC simulations for a
trapped polymer between two repulsive walls \cite{MB98,HG04}. The
plots for ideal chain (exact) and real polymer chain with the EVI
(wide slit) represent the results of our calculations in the
framework of MFT. The first SAW (MC) are the results obtained by
\cite{MB98}. The curves depicted as RW (MC) and the second SAW (MC)
are the results obtained in the narrow slit limit by \cite{HG04} for
random walks and self avoiding walks.} \label{fig:gamma-comp}
\end{center}
\end{figure}

 In the case of narrow slit with $y<<1$ the asymptotic solution for
the depletion force in the case of two repulsive walls and for the
case of one repulsive and the other one inert wall simply becomes
$\Gamma_{DD(DN),narr}(y){\approx}-1$ (see \cite{RU09}) for ideal
polymer chains. These results can be understood phenomenologically.
In our units the quantities ${\Theta}$ and ${\Gamma}$ were
normalized to the overall polymer density $n_p$. So, the above
results simply indicate that the force is entirely induced by free
chains surrounding the slit, or, in other words, by the full bulk
osmotic pressure from the outside of the slit. No chain has remained
in the slit. It is reasonable in the case of repulsive walls in the
limit of narrow slits. The above mentioned arguments were used in
order to obtain the leading contributions to the depletion effect as
$y\to 0$. We can state that in the case of very narrow slits
polymers would pay a very high entropy to stay in the slit or even
enter it. It is due to the fact that the phase space containing all
possible conformations is essentially reduced by the squeezing
confinement to the size $\frac{d-1}{d}$\,times its original size
(for an unconfined chain). Therefore, as it was shown in
\cite{RU09}, the ratio of partition function of polymer chain in the
slit and the free chain partition function vanishes strongly as
$y\to 0$. The advantage of the proposed procedure is that no
expansion is necessary in this case of narrow slit region and it
should be equally valid for polymers with the EVI.

Besides, in \cite{RU09,RUIOP09} we obtained very good agreement of
our analytical results for the reduced canonical force $LK/k_{B}T$
and the results obtained by Hsu and Grasberger \cite{HG04} based on
the lattice MC algorithm on a regular cubic lattice in three
dimensions, as it is possible to see on Figure 3$(b)$. In general
the canonical free energy can be obtained via the Legendre transform
from the grand canonical one in the thermodynamic limit $({\tilde
N}, V \to \infty)$ (for details see \cite{RU09}).

\section{Ideal and real polymer chains in a solution of mesoscopic spherical particles}
\renewcommand{\theenumi}{\arabic{enumi}}
The task of a great importance is investigation of a dilute solution
of ideal and real polymer chains with the EVI in a good solvent
immersed in the solution of mesoscopic colloidal particles or
nanoparticles. The interaction of long flexible nonadsorbing linear
polymer chains with mesoscopic particles of big or small size with
different shape was the subject of investigation in a series of
papers \cite{BEShH99,HED99}. In our papers
\cite{RU09,RUIOP09,UKChR17} we focused attention on the
investigation of the depletion interaction potential and the
depletion force for two cases: $(1)$ between a big spherical
colloidal particle of radius $R$ and the wall; $(2)$ between two
colloidal particles of big sizes with different radii $R_{1}$ and
$R_{2}$ and different adsorbing or repelling properties with respect
to polymers in the solution. The interaction of a dilute solution of
ideal and real linear polymer chains with particles and walls is
implemented by the corresponding boundary condition. The difference
between the forces with and without the particle (or particles)
yields the depletion interaction potential of the particle with the
wall (or between two particles). Taking into account the Derjaguin
approximation \cite{D34}, which describes the sphere of the big
colloidal particle of the radius $R$ (with $R>>L$ and $R>>R_{x}$) by
a superposition of immersed plates (see Figure 4(a)) with local
distance $h(\rho)=a+R-\sqrt{R^{2}-\rho^{2}}$ from the wall, where
$a$ - is the nearest distance from the particle to the wall and
$\rho$ - is the width of the fringe itself we performed in
\cite{RU09,UKChR17} the calculations of the depletion interaction
potential ${\Phi}({\tilde{y}})/n_{p}k_{B}T$ and the depletion force
$-\frac{1}{n_{p}k_{B}T}\frac{d{\Phi}({\tilde{y}})}{d{\tilde{y}}}$
between the colloidal particle and the wall in a dilute solution of
linear polymer chains. We also used some modifications of the
Derjaguin approximation for the calculation of the depletion
interaction potential and the depletion force in the case of two big
spherical colloidal particles with different radii $R_{1}\neq R_{2}$
when $R_{i}>>L$ and $R_{i}>>R_{x}$, $i=1,2$.  In this case we took
into account that the distance $h(\rho)$ is equal to:
$h(\rho)=a+R_{1}-\sqrt{R_{1}^{2}-\rho^{2}}+R_{2}-\sqrt{R_{2}^{2}-\rho^{2}}$,
where $a$ - is the nearest distance from the particle to other
particle (see Figure 4(b)). Thus, the depletion interaction
potential ${\Phi}({\tilde{y}})/n_{p}k_{B}T$ can be written in the
form: $2\pi{\tilde R} R^{2}_{x}\int_{\tilde{y}}^{\infty}dy\Theta(y)$
with ${\tilde{y}}=a/R_{x}$ where ${\tilde R}=R$ for the case of the
big spherical colloidal particle of radius $R$ near the wall and
${\tilde R}=R_{1}R_{2}/(R_{1}+R_{2})$ for the case of two big
spherical colloidal particles with different radii $R_{1}\neq
R_{2}$.
\begin{figure}[ht]
        \begin{center}
\includegraphics[width=5.0cm]{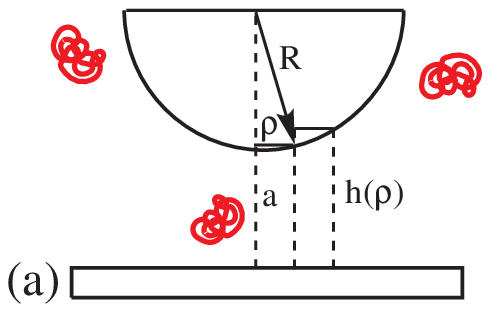}\hspace*{0.3cm}
\includegraphics[width=5.0cm]{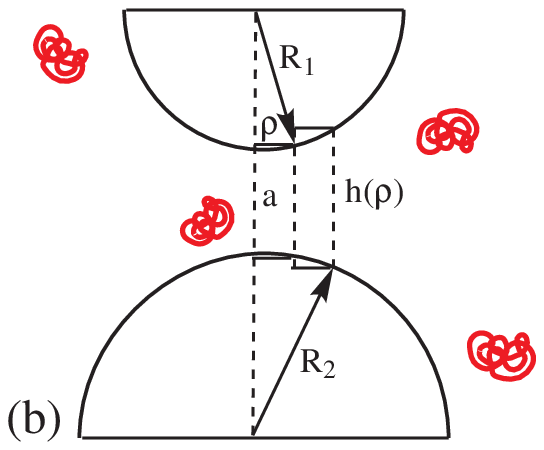}
\caption{The schematic representation of the Derjaguin approximation
(see \cite{D34})\label{SWSS_drawing}}
        \end{center}
    \end{figure}
 Taking into account the results for the depletion interaction
potential ${\Phi}({\tilde{y}})/n_{p}k_{B}T$ (see
\cite{RU09,RUIOP09}) we performed calculation of the depletion force
for certain b.c.'s. The depletion force between the big spherical
colloidal particle and the wall (or between two big spherical
colloidal particles) has the following form \cite{UHK16,UKChR17}:
\be
     - \frac{1}{n_pk_BT}\frac{d\Phi^{id}(\tilde{y})}{d \tilde{y}}
      = 2 \pi
     R_{x}^{2} {\tilde R}
     \vartheta^{id}(\tilde{y}),\label{forcepwpp}\ee
where the results for the corresponding scaling functions
$\vartheta^{id}(\tilde{y})$ for the case of a dilute solution of
ideal polymer chains are presented in Figures 5(a)-5(c) by blue
lines with blue circles, respectively. It should be mentioned that
in a similar way we performed calculations for the depletion force
between the mesoscopic colloidal particle of big size and the wall
(or between two big spherical colloidal particles of different
radii) induced by a dilute solution of real polymer chains with the
EVI in a good solvent.
\begin{figure}[h]
\begin{center}
\includegraphics[width=5.0cm]{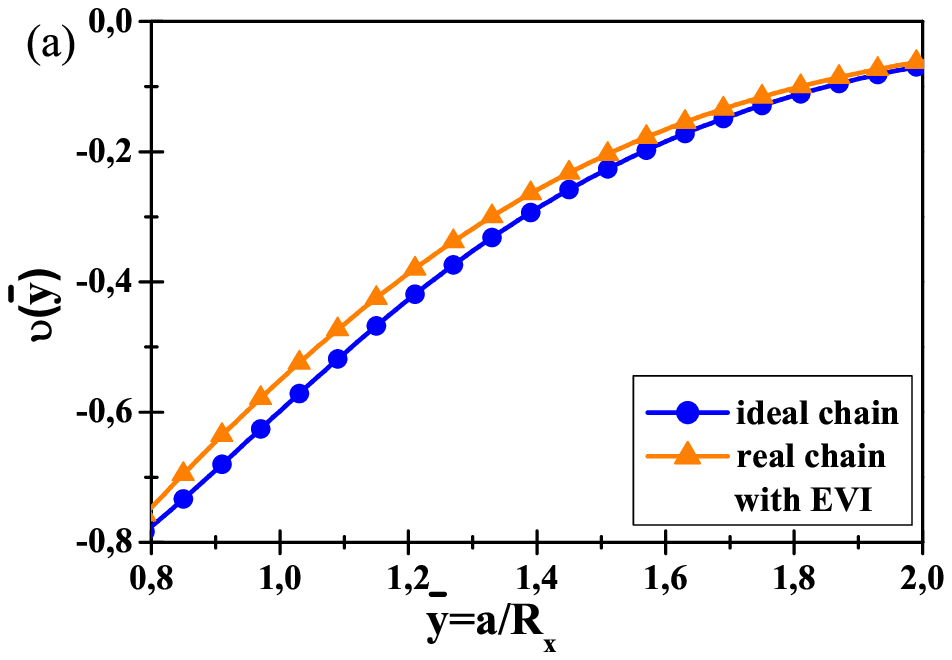}
\includegraphics[width=5.0cm]{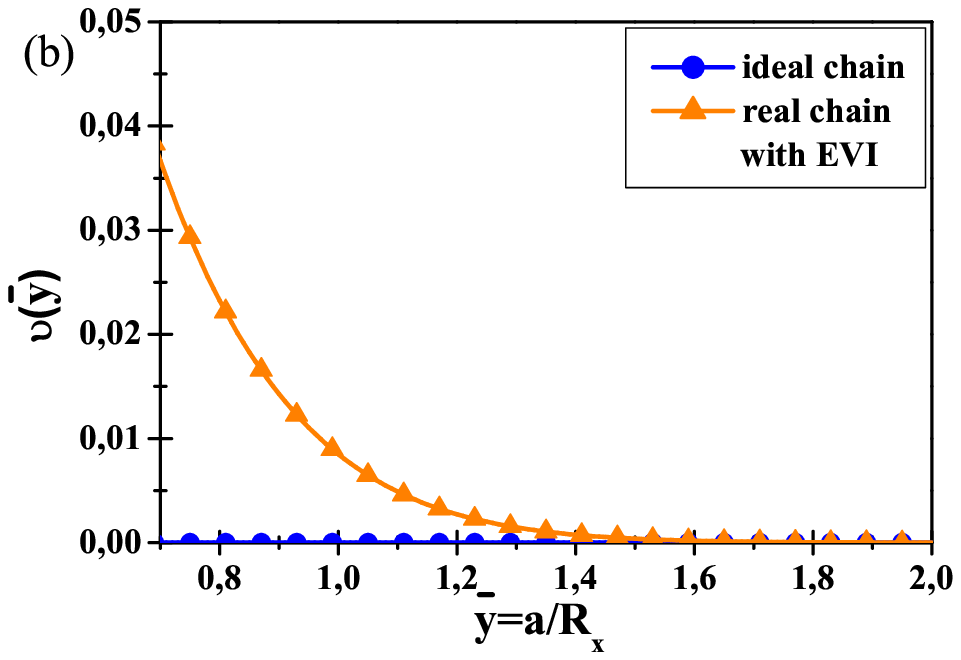}
\includegraphics[width=5.0cm]{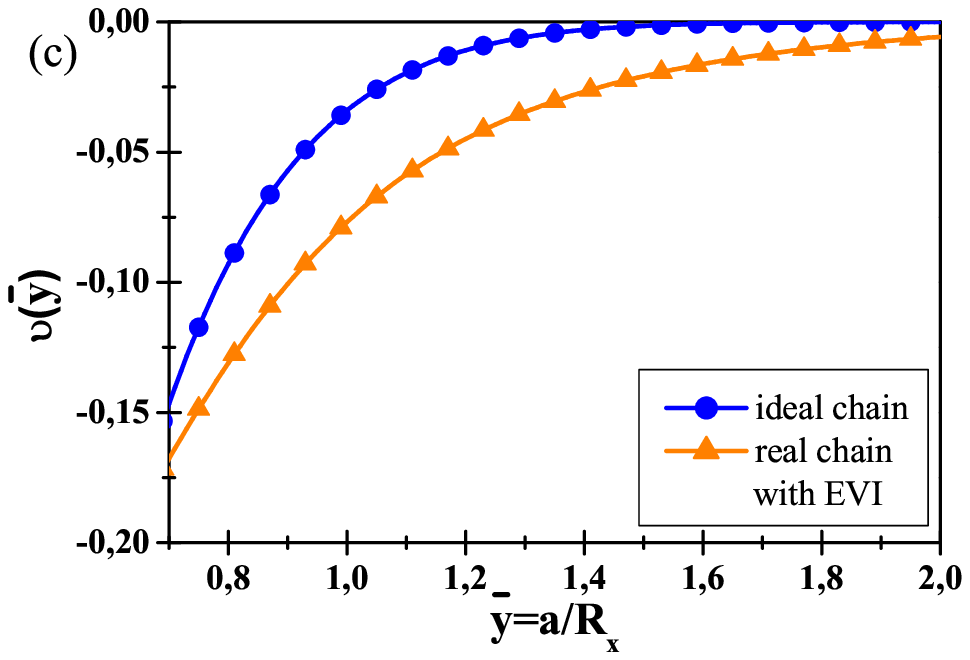}
\caption{The dimensionless scaling functions of
$\vartheta(\tilde{y})$ for a dilute solution of linear polymers
immersed between the big spherical colloidal particle and the wall
(or between two big spherical colloidal particles) for the case of
$(a)$ D-D b.c.; $(b)$ N-N b.c.; $(c)$ D-N b.c.}
\label{fig:gamma_sph}
\end{center}
\end{figure}
The obtained results for the corresponding dimensionless scaling
functions $\vartheta^{EVI}(\tilde{y})$ of the depletion force in the
case of real polymer chains with the EVI in a good solvent immersed
in the solution of big spherical colloidal particles are presented
in Figures 5(a)-5(c) by orange lines with orange triangles,
respectively (see also \cite{UKChR17}). It should be mentioned that
in this situation we also distinguished two cases at the calculation
of the depletion interaction potential and the depletion force:
$(1)$ between a big spherical colloidal particle of radius $R$ and
the wall; $(2)$ between two colloidal particles of big sizes with
different radii $R_{1}$ and $R_{2}$. Besides, we took into account
different adsorbing or repelling properties between the mesoscopic
colloidal particle and the wall (or between two big spherical
colloidal particles) with respect to polymers in the solution which
were implemented by taking into consideration the different boundary
conditions.

As it is possible to see from the obtained results (see
Eq.\ref{forcepwpp}) the absolute value of the depletion force
between the mesoscopic spherical colloidal particle and the wall is
bigger than for the case of two colloidal particles. This feature is
universal and does not depend on the boundary condition type. In the
case when two colloidal particles have the same radius, the
corresponding depletion force is twice smaller than for the case of
the particle near the wall.

The improvement of the experimental technique allowed to measure
with high accuracy the depletion force between a wall and a single
colloidal particle \cite{OSO97,RBL98,VCLY98}. In one of our previous
papers \cite{RU09} the good qualitative agreement between analytical
results obtained in the framework of the massive field theory
approach at fixed space dimensions $d=3$ and the experimental
results obtained by Rudhardt, Bechinger and Leiderer \cite{RBL98} by
means of total internal reflection microscopy techniques (TIRM) for
the depletion interaction potential between the spherical colloidal
particle of big radius $R=1.5\mu m$ and the wall of the container
immersed in the dilute solution of nonionic linear polymer chains
were obtained in the case when the radius of gyration was used as
freely adjustable parameter (we assumed that $R^{fit}_{g}=0.13\mu
m$) (see Figure 6).

\begin{figure}[ht!]
\begin{center}
\includegraphics[width=8.0cm]{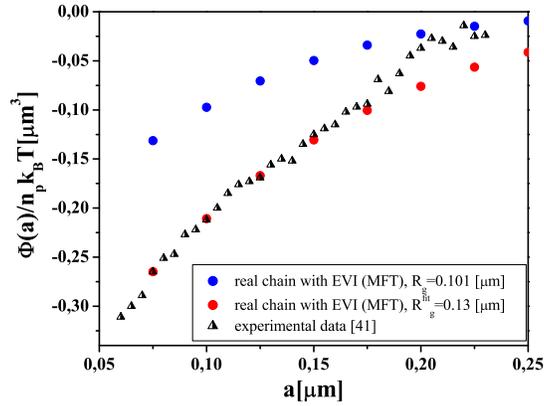}
\caption{Comparison of the massive field theory (MFT) results
obtained for the the depletion interaction potential
$\frac{\Phi(a)}{n_{p}k_{B}T}$ with the experimental results obtained
by means of total internal reflection microscopy techniques
\cite{RBL98}, where $a$ - is the closest distance between the
particle and the wall. In the experiment \cite{RBL98} the radius of
colloidal particle was assumed to be $R=1.5 \mu m$ and the radius of
gyration has been determined by means of small angle scattering
experiments ($R_{g}=0.101\mu m$).}\label{Fig:1}
\end{center}
\end{figure}

\section{The universal density-force
relation and the monomer density profiles}
\renewcommand{\theenumi}{\arabic{enumi}}

One of the most remarkable theoretical predictions proposed by
Joanny, Leibler and de Gennes \cite{JLG} states that the monomer
density profile of a dilute solution of long flexible linear polymer
chains bounded by a planar repulsive wall has a depletion region of
mesoscopic width of order of the coil size $R_{x}\approx N^{\nu}$
(where $N$ is the number of monomers per chain), and for the
distances $z$ from the wall that are small compared to this width
but bigger than microscopic lengths of monomer size $\tilde{l}$  the
profile increases according to the law
$$\rho(z)\sim z^{\frac{1}{\nu}}$$ with Flory exponent $\nu$.
The critical exponent $\nu$ is equal $1/2$ for ideal polymer chains
and $\nu\approx0.588$ for real polymer chains with the EVI in a good
solvent. As it was mentioned in \cite{JLG}, the monomer density
close to the wall is proportional to the force per unit area which
the polymer solution exerts on the wall. In a paper \cite{E97} a
complete quantitative expression for the universal density-force
relation was obtained for the first time by Eisenriegler on the
basis of $\epsilon$- expansion up to first order. The validity of
the above mentioned density-force relation for the different cases
of a single polymer chain with one end (or both ends) fixed in the
half space bounded by the wall, a single chain trapped in the slit
geometry of two parallel walls, for the case of dilute and
semi-dilute solution of free polymer chains in a half space and for
the case of polymer chain in a half space containing a mesoscopic
particle of arbitrary shape was discussed.

The interaction of long flexible nonadsorbing linear polymer chains
with mesoscopic colloidal particles of big and small size and
different shape was the subject a series of papers
\cite{BEShH99,HED99}. The above mentioned universal density-force
relation was verified by simulation methods using an off-lattice
bead-spring model of a polymer chain trapped between two parallel
repulsive walls \cite{MB98} and by the lattice Monte Carlo (MC)
algorithm on a regular cubic lattice in $d=3$ dimensions
\cite{HG04}. In a series of our papers \cite{U11,Ujml11} the
universal density-force relation was analyzed by analogy as it was
proposed by Eisenriegler \cite{E97} and the corresponding universal
amplitude ratio was obtained in the framework of the massive field
theory approach directly in fixed space dimensions $d=3$. It allowed
us to obtain in a series of our papers \cite{U11,Ujml11} the monomer
density profiles of ideal and real linear polymer chains with the
EVI in a good solvent immersed in a slit geometry of two parallel
repulsive walls, one repulsive and the other one inert wall. The
obtained in a series of our papers \cite{U11,Ujml11} results
indicate that in the case of two repulsive walls the maximum of the
layer monomer density is situated in the middle of the slit at
$L/2$. In the case of one repulsive and the other one inert wall the
maximum of the layer monomer densities is near the distant inert
wall. Besides, the monomer density profiles of a dilute polymer
solution confined in a semi-infinite space containing the mesoscopic
spherical colloidal particle of big size with different adsorbing or
repelling properties in respect for polymers were calculated (see
\cite{U11,Ujml11}).

\section{Conclusions}
\renewcommand{\theenumi}{\arabic{enumi}}

Summarizing the obtained in a series of papers
\cite{SKB09,BRSSU13,U14,Umac14} results we should mention that the
performed investigations have important practical applications for
better understanding of the elastic properties of the individual
macromolecules, networks, gels and brush layers. The obtained in the
framework of the massive field theory approach results for the
stretching force are in good agreement with the experimental results
based on AFM measurements \cite{HS01,NHNG06} as it was shown in
\cite{U14} and with the results obtained in the framework of DFT and
MC methods \cite{BRSSU13} for the region of small applied forces
when deformation of polymer chain is not bigger than linear
extension of polymer chain in this direction. Besides, the better
agreement between the analytical results and the results obtained in
the framework of DFT and MC methods is observed in the case, when
the number of monomers increases and polymer chain becomes longer.
In the case of strong deformations of polymer chain the Gaussian
distribution does not give good results for the stretching force and
in this case the calculations should be performed with taking into
account Langevin distribution function.

Besides, the results obtained in
\cite{E97,ShHKD01,RU09,RUIOP09,U11,Ujml11,UKChR17} indicate that for
the entropic reasons, the polymer chains avoid the space between two
repulsive walls in a slit geometry or between two close colloidal
particles with repelling properties in respect to polymers. In
confined dilute polymer solutions  are present the depletion regions
near the confining walls or mesoscopic particles due to an
additional amount of entropic energy for polymers. It leads to an
unbalanced pressure the outside which pushes the two repulsive walls
or two such colloidal particles towards each other. As a result the
attractive depletion force arises between two repulsive walls or two
close colloidal particles with repelling properties.

Completely different behavior is observed for the case of a dilute
polymer solution of long-flexible linear polymer chains confined in
a slit of two inert walls (the walls where the adsorption threshold
takes place). One of the remarkable results which we obtained in
\cite{RU09,RUIOP09} shows that the depletion force in the case of
two inert walls  becomes repulsive for the case of linear real
polymer chains with the EVI in a good solvent. This result is very
important from practical point of view, because it means that in
such systems we observe reduction of the static friction and as a
result such systems can be used for producing of new types of nano-
and micro-electromechanical devices.

The performed in a series of papers
\cite{E97,BEShH99,HED99,RU09,RUIOP09,UKChR17} investigations for
long flexible linear polymer chains immersed in a solution of
mesoscopic colloidal particles of big or small size with different
shape indicate that focusing on such systems leads to universal
results which are independent of microscopic details and depend only
on shapes of particles, adsorbing or repelling properties of the
confining walls in respect to polymers and ratios of three
characteristic lengths of the system such as the radius of the
colloidal particle, the polymer size, the distance between the
colloidal particle and the wall or between two mesoscopic colloidal
particles, respectively.

 In general, the analysis of the obtained in the framework of the
massive field theory approach at fixed space dimensions $d=3$
results for the surface critical exponents \cite{U06}, the monomer
density profiles \cite{U11,Ujml11}, the stretching force in the case
of anchored to the surface ideal or real linear polymer chain with
the EVI in a good solvent \cite{U14,Umac14}, as well as the results
of calculations of the dimensionless depletion interaction potential
and the depletion force which arise in a dilute polymer solution
immersed between two parallel walls in a slit \cite{RU09,RUIOP09} or
between colloidal particle and the wall (or between two colloidal
particles) \cite{RU09,UKChR17} indicates about very good agreement
between the results of this analytical method and other theoretical
methods \cite{EKB82,E97,SKB09,ShHKD01}. Besides, as it was possible
to observe in a series of our papers
\cite{RU09,RUIOP09,U11,Ujml11,U14,Umac14} there is a very good
agreement between our analytical results and numerical results based
on MC calculations or DFT approach \cite{HG04,BRSSU13}, and the
experimental results obtained by different techniques such as AFM
experiment \cite{HS01,NHNG06} and total internal reflection
microscopy (TIRM) \cite{RBL98}. All this confirm that the massive
field theory approach at fixed space dimensions $d<4$ is a promising
candidate for description of critical behavior of other complex
systems in confined geometries.

\section{Author Contributions}
\renewcommand{\theenumi}{\arabic{enumi}}

Z.Usatenko performed calculations, write the review paper and
prepared part of figures. K.S. Danel search for the literature,
prepared reference list and part of figures.

\section{Conflict of Interests}
\renewcommand{\theenumi}{\arabic{enumi}}

The authors declare no conflict of interests.

\end{flushleft}

\begin{thebibliography}{99}
\singlespacing

\bibitem{NLA07} Neumann, K.C.; Lionnet, T.; Allemand, J.F. Single-Molecule Micromanipulation Techniques.
{\it Annu. Rev. Mater. Res.} {\bf 2007}, {\it 37}, 33-67, DOI:
10.1146/annurev.matsci.37.052506.084336. Available online:
http://www.annualreviews.org/doi/pdf/10.1146/annurev.matsci.37.052506.084336
(accessed on 11 January 2007).

\bibitem{NN08} Neuman, K.C.; Nagy, A. Single-molecule force spectroscopy: Optical tweezers, magnetic tweezers
and atomic force microscopy. {\it Nat. Methods} {\bf 2008}, {\it 5},
491-505, DOI: 10.1038/nmeth.1218. Available online:
https://www.nature.com/articles/nmeth.1218 (accessed on 29 May
2008).

\bibitem{WR02} Williams, M.C.; Rouzina, I.  Force spectroscopy of single DNA and RNA molecules. {\it Curr. Opin. Struct.
Biol.} {\bf 2002}, {\it 12}, 330-336,
DOI:10.1016/S0959-440X(02)00340-8. Available online:
http://www.sciencedirect.com/science/article/pii/S0959440X02003408?via\%
3Dihub (accessed on 17 July 2002).

\bibitem{HS01} Hugel, T.; Seitz, M. The Study of Molecular Interactions by AFM Force
Spectroscopy.{\it Macromol. Rapid Commun.} {\bf 2001}, {\it 22},
989-1016,
DOI:10.1002/1521-3927(20010901)22:13<989::AID-MARC989>3.0.CO;2-D.
Available online:
http://onlinelibrary.wiley.com/doi/10.1002/1521-3927(20010901)22:13\%
3C989::AID-MARC989\%3E3.0.CO;2-D/abstract (accessed on 1 September
2001).

\bibitem{HHCMSG02} Hugel, T.; Holland, N.B.; Cattani, A.; Moroder,
L.; Seitz, M.; Gaub, H.E. Single-molecule optomechanical cycle. {\it
Science} {\bf 2002}, {\it 296}, 1103-1106,
DOI:10.1126/science.1069856. Available online:
http://science.sciencemag.org/content/296/5570/1103 (accessed on 10
May 2002).

\bibitem{HHNCROMSG03} Holland, N.B.; Hugel, T.; Neuert, G.;
Cattani-Scholz, A.; Renner, C.; Oesterhelt, D.; Moroder, L.; Seitz,
M.; Gaub, H.E. Single Molecule Force Spectroscopy of Azobenzene
Polymers: Switching Elasticity of Single Photochromic
Macromolecules. {\it Macromolecules} {\bf 2003}, {\it 36},
2015-2023, DOI:10.1021/ma021139s. Available online:
http://pubs.acs.org/doi/abs/10.1021/ma021139s (accessed on February
25 2003).

\bibitem{UHSGShDXP02} Urry, D.W.; Hugel, T.; Seitz, M.; Gaub, H.E.;
Sheiba, L.; Dea, J.; Xu, J.; Parker, T. Elastin: a representative
ideal protein elastomer. {\it Philos. Trans. R. Soc. London, Ser. B:
Biol. Sci.} {\bf 2002}, {\it 357}, 169-184,
DOI:10.1098/rstb.2001.1023. Available online:
http://rstb.royalsocietypublishing.org/content/357/1418/169
(accessed on 28 February 2002).

\bibitem{ROHG97} Rief, M.; Oesterhelt, F.; Heymann, B.; Gaub, H.E.
Single Molecule Force Spectroscopy on Polysaccharides by Atomic
Force Microscopy. {\it Science} {\bf 1997}, {\it 275}, 1295-1297,
DOI:10.1126/science.275.5304.1295. Available online:
https://www.ncbi.nlm.nih.gov/pubmed/9036852 (accessed on 28 February
1997).

\bibitem{FBJSG04} Friedsam, C.; Becares, A.D.; Jonas, U.; Seitz, M.;
Gaub, H.E. Adsorption of polyacrylic acid on self-assembled
monolayers investigated by single-molecule force spectroscopy. {\it
New J. Phys.} {\bf 2004}, {\it 6}, DOI:10.1088/1367-2630/6/1/009.
Available online:
http://iopscience.iop.org/article/10.1088/1367-2630/6/1/009/pdf
(accessed on 30 January 2004).

\bibitem{FSG04} Friedsam, C.; Seitz, M.; Gaub, H.E. Investigation of
polyelectrolyte desorption by single molecule force spectroscopy.
{\it J. Phys.: Condens. Matter} {\bf 2004}, {\it 16}, S2369-S2382,
DOI:10.1088/0953-8984/16/26/010. Available online:
http://iopscience.iop.org/article/10.1088/0953-8984/16/26/010/meta
(accessed on 18 June 2004).

\bibitem{SFJHG03} Seitz, M.; Friedsam, C.; Jostl, W.; Hugel, T.;
Gaub, H.E. Probing Solid Surfaces with Single Polymers.{\it
ChemPhysChem} {\bf 2003}, {\it 4}, 986-990, DOI:
10.1002/cphc.200300760. Available online:
http://onlinelibrary.wiley.com/doi/10.1002/cphc.200300760/abstract
(accessed on 10 September 2003).

\bibitem{HRSGN05} Hugel, T.; Rief, M.; Seitz, M.; Gaub, H.E.; Netz,
R.R. Highly Stretched Single Polymers: Atomic-Force-Microscope
Experiments Versus Ab-Initio Theory. {\it Phys. Rev. Lett.} {\bf
2005}, {\it 94}, 048301, DOI:10.1103/PhysRevLett.94.048301.
Available online:
https://journals.aps.org/prl/abstract/10.1103/PhysRevLett.94.048301
(accessed on 4 February 2005).

\bibitem{KEG06} Kuhner, F.; Erdmann, M.; Gaub, H.E. Scaling exponent and Kuhn length of pinned
polymers by single molecule force spectroscopy. {\it Phys. Rev.
Lett.} {\bf 2006}, {\it 97}, 21831, DOI:
10.1103/PhysRevLett.97.218301. Available online:
https://journals.aps.org/prl/abstract/10.1103/PhysRevLett.97.218301
(accessed on 20 November 2006).

\bibitem{NHNG06} Neuert, G.; Hugel, T.; Netz, R.R.; Gaub, H.E.
Elastisity of Poly(azobenzene-peptides). {\it Macromolecules} {\bf
2006}, {\it 39}, 789-797, DOI:10.1021/ma051622d. Available online:
http://pubs.acs.org/doi/abs/10.1021/ma051622d (accessed on 15
December 2005).

\bibitem{KPG08} Kufer, S.K.; Puchner, E.M.; Gumpp, H.; Liedl, T.; Gaub H.E. Single-molecule cut-and-paste surface assembly
{\it Science} {\bf 2008}, {\it 319}, 594,
DOI:10.1126/science.1151424. Available online:
https://www.ncbi.nlm.nih.gov/pubmed/18239119 (accessed on 01
February 2008).

\bibitem{ORG99} Oesterhelt, F.; Rief, M.; Gaub, H.E. Single molecule force spectroscopy by AFM
indicates helical structure of poly(ethylene-glycol) in water. {\it
New J. Phys.} {\bf 1999}, {\it 1}, 6.1-6.11, DOI:PII.
S1367-2630(99)99121-8. Available online:
http://iopscience.iop.org/article/10.1088/1367-2630/1/1/006/pdf
(accessed on 24 March 1999).

\bibitem{LNK03} Livadaru, L.; Netz, R.R.; Kreuzer, H.J. Stretching Response of Discrete Semiflexible Polymers. {\it
Macromoecules} {\bf 2003}, {\it 36}, 3732, DOI:10.1021/ma020751g.
Available online: http://pubs.acs.org/doi/abs/10.1021/ma020751g
(accessed on 25 April 2003).

\bibitem{RC03} Rubinstein, M.; Colby, R.H. {\it Polymer Physics}, Oxford
University Press: New York, USA, 2003; ISBN 978-0198520597.

\bibitem{OOPEGM00} Oesterhelt, F.; Oesterhelt, D.; Pfeiffer, M.;
Engel, A.; Gaub, H.E.; Muller, D.J. Unfolding pathways of individual
bacteriorhodopsins. {\it Science} {\bf 2000}, {\it 288}, 143, DOI:
10.1126/science.288.5463.143. Available online:
http://science.sciencemag.org/content/288/5463/143.long (accessed on
7 Apr 2000).

\bibitem{KSGB97} Kellermayer, M.S.Z.; Smith, S.B.; Granzier, H.L.;
Bustamante, C. Folding-unfolding transitions in single titin
molecules characterized with laser tweezers. {\it Science} {\bf
1997}, {\it 276}, 1112, DOI:10.1126/science.276.5315.1112. Available
online: http://science.sciencemag.org/content/276/5315/1112.long
(accessed on 16 May 1997).

\bibitem{KWG99} Kreuzer, H.J.; Wang, R.L.C.; Grunze, M. Effect of stretching on the molecular
conformation of oligo (ethylene oxide): A theoretical study.{\it New
J. Phys.} {\bf 1999}, {\it 1}, 21.1-21.16, DOI:PII.
S1367-2630(99)06011-5. Available online:
http://iopscience.iop.org/article/10.1088/1367-2630/1/1/321/pdf
(accessed on 26 November 1999).

\bibitem{B00} Bhattacharjee S.M. Unzipping DNAs: Towards the first step of replication. {\it J. Phys. A} {\bf 2000}, {\it
33}, L423-L428, DOI:10.1088/0305-4470/33/45/101. Available online:
http://iopscience.iop.org/article/10.1088/0305-4470/33/45/101/meta
(accessed on 17 November 2000).

\bibitem{MTM01} Marenduzzo, D.; Trovato, A.; Maritan, A. Phase diagram of force-induced DNA unzipping in exactly solvable models. {\it Phys.
Rev. E} {\bf 2001}, {\it 64}, 031901,
DOI:10.1103/PhysRevE.64.031901. Available online:
https://link.aps.org/doi/10.1103/PhysRevE.64.031901 (accessed on 14
August 2001).

\bibitem{E93} Eisenriegler, E. {\it Polymers Near Surfaces}, World
Scientific Publishing Co. Pte. Ltd.: Singapore, 1993; ISBN
978-981-02-0595-9.

\bibitem{EKB82} Eisenriegler, E.; Kremer, K.; Binder, K. Adsorption of polymer chains at surfaces: Scaling and Monte Carlo analyses. {\it
J. Chem. Phys.} {\bf 1982}, {\it 77}, 6296, DOI:10.1063/1.443835.
Available online: http://aip.scitation.org/doi/abs/10.1063/1.443835
(accessed on June 1998).


\bibitem{WNF87} Wang, Zhen-Gang.; Nemirovsky, A.M.; Freed Karl F. Polymers with excluded volume in
various geometries: Renormalization group methods. {\it J. Chem.
Phys.} {\bf 1987},  {\it 86}, 4266, DOI:10.1063/1.451887. Available
online: https://doi.org/10.1063/1.451887 (accessed on August 1998).

\bibitem{U06} Usatenko, Z. Adsorption of long-flexible polymer chains
at planar surfaces: Scaling analysis and critical exponents. {\it
J.Stat.Mech.} {\bf 2006}, P03009,
DOI:10.1088/1742-5468/2006/03/P03009. Available online:
http://iopscience.iop.org/article/10.1088/1742-5468/2006/03/P03009
(accessed on 13 March 2006).

\bibitem{U11} Usatenko, Z. Monomer density profiles for polymer chains in confined
geometries: massive field theory approach. {\it J. Chem. Phys.} {\bf
2011}, {\it 134}, 024119, DOI:10.1063/1.3529426. Available online:
http://aip.scitation.org/doi/10.1063/1.3529426 (accessed on January
2011).

\bibitem{Ujml11} Usatenko, Z. Monomer density profiles of real polymer chains in
confined geometries. {\it J. Mol. Liq.} {\bf 2011}, {\it 164},
59-65, DOI:10.1016/j.molliq.2011.06.010. Available online:
http://www.sciencedirect.com/science/article/pii/S0167732211002157
(accessed on 21 July 2011).

\bibitem{UC04} Usatenko, Z.; Ciach, A. Critical adsorption of polymers in
a medium with long-range correlated disorder. {\it Phys. Rev. E}
{\bf 2004}, {\it 70}, 051801, DOI:10.1103/PhysRevE.70.051801.
Available online:
https://link.aps.org/doi/10.1103/PhysRevE.70.051801 (accessed on 9
November 2004).


\bibitem{SKB09} Skvortsov, A.M.; Klushin, L.I.; Birshtein, T.M.
Stretching and compression of a macromolecule under different modes
of mechanical manupulations. {\it Polymer Science, Ser.A.} {\bf
2009}, {\it 51}, 469-491, DOI:10.1134/S0965545X09050010. Available
online: https://link.springer.com/article/10.1134/S0965545X09050010
(accessed on 09 June 2009).

\bibitem{BRSSU13} Bor\'owko, M.; R\.zysko, W.; Sokolowski, S.;
Sokolowska, Z.; Usatenko, Z. Stretching tethered polymer chain:
Density functional approach. {\it J. Chem. Phys.} {\bf 2013}, {\it
138}, 204707, DOI:10.1063/1.4807086. Available online:
http://aip.scitation.org/doi/10.1063/1.4807086 (accessed on May
2013)

\bibitem{U14} Usatenko, Z. Stretching of polymer chain anchored
to a  surface: the massive field theory approach. {\it J. Stat.
Mech.: Theory and Experiment} {\bf 2014}, {\it P09015},
DOI:10.1088/1742-5468/2014/09/P09015. Available online:
https://doi.org/10.1088/1742-5468/2014/09/P09015 (accessed on 16
September 2014).

\bibitem{Umac14} Usatenko, Z. Stretching of anchored to the surface
a real polymer chain in a good solvent: the massive field theory
approach. {\it Macromolecular Symposia} {\bf 2014}, {\it 346},
14-21, DOI:10.1002/masy201400029. Available online:
onlinelibrary.wiley.com/doi/10.1002/masy.201400029/full (accessed on
23 December 2014).

\bibitem{Par80} Parisi, G. Field-theoretic approach to second-order phase transitions
in two- and three-dimensional systems.{\it J.Stat.Phys.} {\bf 1980},
{\it 23}, 49-82, DOI:10.1007/BF01014429. Available online:
https://link.springer.com/article/10.1007\%2FBF01014429 (accessed on
July 1980).

\bibitem{DSh98}
Diehl, H.W.; Shpot, M. Massive field-theory approach to surface
critical behaviour in three - dimensional systems. {\it Nucl. Phys.
B} {\bf 1998}, {\it 528}, 595-647,
DOI:10.1016/S0550-3213(98)00489-1. Available online:
https://doi.org/10.1016/S0550-3213(98)00489-1 (accessed on 21
September 1998).

\bibitem{RU09} Romeis, D.; Usatenko, Z. Polymer chains in confined geometries:
Massive field theory approach. {\it Phys.Rev.E} {\bf 2009}, {\it
80}, 041802, DOI:10.1103/PhysRevE.80.041802. Available online:
https://link.aps.org/doi/10.1103/PhysRevE.80.041802 (accessed on 7
October 2009).

\bibitem{RUIOP09} Romeis, D.; Usatenko, Z. Massive field theory approach for polymer
chains in confined geometries. {\it AIP Conference Proceedings,
American Institute of Physics} {\bf 2009}, {\it 1198},  144-155,
DOI:10.1063/1.3284410. Available online:
http://aip.scitation.org/doi/abs/10.1063/1.3284410 (accessed on
December 2009).

\bibitem{ShHKD01} Schlesener, F.; Hanke, A.; Klimpel, R.; Dietrich,
S. Polymer depletion interaction between two parallel repulsive
walls. {\it Phys.Rev.E} {\bf 2001}, {\it 63}, 041803,
DOI:10.1103/PhysRevE.63.041803. Available online:
https://journals.aps.org/pre/abstract/10.1103/PhysRevE.63.041803
(accessed on 27 March 2001).


\bibitem{UKChR17} Usatenko, Z.; Kuterba, P.; Chamati, H.; Romeis, D. Linear
and ring polymers in confined geometries. {\it Eur.Phys. J. Special
Topics} {\bf 2017} {\it 226}, 651-665,
DOI:10.1140/epjst/e2016-60335-0. Available online:
link.springer.com/article/10.1140/epjst/e2016-60335-0 (accessed on
05 April 2017).

\bibitem{RBL98} Rudhardt, D.; Bechinger, C.; Leiderer, P. Direct Measurement of Depletion Potentials
in Mixtures of Colloids and Nonionic Polymers. {\it Phys.Rev.Lett.}
{\bf 1998}, {\it 81}, 1330, DOI:10.1103/PhysRevLett.81.1330.
Available online:
https://journals.aps.org/prl/abstract/10.1103/PhysRevLett.81.1330
(accessed on 10 August 1998).

\bibitem{JLG} Joanny, J.F.; Leibler, L.; de Gennes, P.G. Effects of polymer solutions on colloid
stability. {\it J. Polym. Sci., Polym. Phys. Ed.} {\bf 1979}, {\it
17}, 1073-1084, DOI:10.1002/pol.1979.180170615. Available online:
http://onlinelibrary.wiley.com/doi/10.1002/pol.1979.180170615/abstract
(accessed on 11 March 2003).

\bibitem{E97} Eisenriegler, E. Universal density-force relations for polymers near a repulsive wall.
{\it Phys. Rev. E} {\bf 1997}, {\it 55}, 3116-3123,
DOI:10.1103/PhysRevE.55.3116. Available online:
https://journals.aps.org/pre/abstract/10.1103/PhysRevE.55.3116
(accessed on 1 March 1997).

\bibitem{D34} Derjaguin, B.V. Untersuchungen ìber die Reibung und Adh¤sion, IV. {\it Kolloid-Z.} {\bf 1934}, {\it 69}, 155-164,
DOI:10.1007/BF01433225. Available online:
https://link.springer.com/article/10.1007\%2FBF01433225 (accessed on
19 June 1934).

\bibitem{BEShH99} Bringer, A.; Eisenriegler, E.; Schlesener, F.; Hanke, A. Polymer depletion interaction between a particle and a wall.
{\it Eur. Phys.J. B} {\bf 1999}, {\it 11}, 101-119,
DOI:10.1007/s100510050921. Available online:
https://epjb.epj.org/articles/epjb/abs/1999/17/b8812/b8812.html
(accessed on 15 September 1999).

\bibitem{HED99} Hanke, A.; Eisenriegler, E.; Dietrich, S. Polymer depletion effects near mesoscopic particles. {\it
Phys.Rev.E} {\bf 1999}, {\it 59}, 6853-6878,
DOI:10.1103/PhysRevE.59.6853. Available online:
https://journals.aps.org/pre/abstract/10.1103/PhysRevE.59.6853
(accessed on 1 June 1999).


\bibitem{AO54} Asakura S.; Oosawa, F. On Interaction between Two Bodies Immersed in a Solution of Macromolecules.
{\it J.Chem.Phys.} {\bf 1954}, {\it 22}, 1255-1256,
DOI:10.1063/1.1740347. Available online:
http://aip.scitation.org/doi/10.1063/1.1740347 (accessed on December
2004).

\bibitem{AO58} Asakura S.; Oosawa, F. Interaction between particles suspended in solutions of macromolecules.
{\it J.Polym.Sci.} {\bf 1958}, {\it 33}, 183-192,
DOI:10.1002/pol.1958.1203312618. Available online:
http://onlinelibrary.wiley.com/doi/10.1002/pol.1958.1203312618/abstract
(accessed on 10 March 2003).

\bibitem{O96} Odijk, T. Protein-Macromolecule Interactions. {\it Macromolecules} {\bf 1996}, {\it 29}, 1842-1843,
DOI:10.1021/ma951467a. Available online:
http://pubs.acs.org/doi/abs/10.1021/ma951467a?journalCode=mamobx
(accessed on 26 February 1996).

\bibitem{O97} Odijk, T. Many-body depletion interactions among protein spheres in a semidilute polymer solution.
 {\it J.Chem.Phys.} {bf 1997}, {\it 106}, 3402-3407,
 DOI:10.1063/1.473069. Available online:http://aip.scitation.org/doi/0.1063/1.473069 (accessed
 on August 1998).

\bibitem{MB98} Milchev A.; Binder, K. A polymer chain trapped between two parallel repulsive walls:
A Monte-Carlo test of scaling behavior. {\it Eur. Phys. J. B} {\bf
1998}, {\it 3}, 477-484, DOI:10.1007/s100510050338.
 Available online: https://link.springer.com/article/10.1007\%2Fs100510050338?LI=true (accessed on 15 June 1998).


\bibitem{HG04} Hsu, H.-P.; Grasberger, P.  Polymers confined between two parallel plane walls. {\it J.Chem.Phys.} {\bf 2004}, {\it 120},
2034-2042, DOI:10.1063/1.1636454. Available online:
http://aip.scitation.org/doi/abs/10.1063/1.1636454 (accessed on
January 2004).

\bibitem{Z89} Zinn-Justin, J. {\it Euclidean Field Theory and Critical
Phenomena}; Oxford Univ. Press: New York, USA, 1989; ISBN
9708198518730.

\bibitem{SD89} Schloms, R.; Dohm, V. Minimal renormalization without $\epsilon$-expansion:
Critical behavior in three dimensions. {\it Nucl. Phys. B} {\bf
1989}, {\it 328}, 639-663, DOI:10.1016/0550-3213(89)90223-X.
Available online:
http://www.sciencedirect.com/science/article/pii/055032138990223X
(accessed on December 1989).

\bibitem{S73} Symanzik, K. {\it Lett. Nuovo Cimento} {\bf 1973}, {\it 8},
771.

\bibitem{BD82} Berg{\`e}re, M.C.; David, F. Nonanalyticity of the perturbative expansion
for super-renormalizable massless field theories. {\it Ann. Phys.
(N.Y.)} {\bf 1982}, {\it 142}, 416-447,
DOI:10.1016/0003-4916(82)90078-1. Available online:
http://www.sciencedirect.com/science/article/pii/0003491682900781
(accessed on September 1982).

\bibitem{DD81} Diehl, H.W.; Dietrich, S. Field-theoretical approach to static critical phenomena in semi-infinite systems.
 {\it Z. Phys. B} {\bf 1981}, {\it 42}, 65-86,
DOI:10.1007/BF01298293. Available online:
https://link.springer.com/article/10.1007\%2FBF01298293 (accessed on
March 1981).

\bibitem{UH17} Usatenko, Z.; Halun, J. Ring polymer chains in confined geometries: the
 massive field theory approach. {\it J. Stat. Mech.: Theory and Experiment} {\bf 2017}, {\it P013303},
 DOI:10.1088/1742-5468/aa5285. Available online: iopscience.iop.org/article/10.1088/1742-5468/aa5285
 (accessed on 27 January 2017).

\bibitem{UHK16} Usatenko, Z.; Halun, J.; Kuterba, P. Ring polymers in confined geometries.
{\it Condensed Matter Physics} {\bf 2016}, {\it 19}, 43602,
DOI:10.5488/CMP.19.43602. Available online:
http://www.icmp.lviv.ua/journal (accessed on 6 September 2016).

\bibitem{UKChH17} Usatenko, Z.; Kuterba, P.; Chamati, H.; Halun, J. Investigation of ring polymers in
confined geometries. {\it Journal of Physics: Conference Series}
{\bf 2017}, {\it 794}, 012002, DOI:10.1088/1742-6596/794/1/12002.
Available online:
iopscience.iop.org/article/10.1088/1742-6596/794/1/012002 (accessed
on 22 February 2017).

\bibitem{deGennes} de Gennes, P.-G. Exponents for the excluded volume problem as derived by the Wilson method.
{\it Phys.Lett.A} {\bf 1972}, {\it 38}, 339-340,
DOI:10.1016/0375-9601(72)90149-1. Available online:
http://www.sciencedirect.com/science/article/pii/0375960172901491?via\%3Dihub
(accessed on 28 February 1972).

\bibitem{deGennes79} de Gennes, P.-G. {\it Scaling Concepts in Polymer Physics}; Cornell University
Press: Ithaca, New York, 1979; ISBN 978-0-8014-1203-5.

\bibitem{CJ90} des Cloizeaux, J.; Jannink, G. {\it Polymers in
Solution}; Clarendon Press, Oxford, 1990; ISBN 9780199588930.

\bibitem{B83} Binder, K. Critical behavior at surfaces. In {\it Phase Transitions and Critical
Phenomena}; Domb, C., Lebowitz, J.L., Eds.; Academic Press: London,
1983; Volume 8, pp.1-144; ISBN 10:0122203089.

\bibitem{D86} Diehl, H.W. Field-theoretical Approach to Critical Behaviour
at Surfaces. In {\it Phase Transitions and Critical Phenomena};
Domb, C., Lebowitz, J.L., Eds.; Academic Press: London, 1986; Volume
10, pp.76- 267; ISBN 0-12-220310-0.

\bibitem{deGennes76} de Gennes, P.G. Scaling theory of polymer adsorption. {\it J.Phys.(Paris)} {\bf 1976}, {\it 37},
1445-1452, DOI:10.1051/jphys:0197600370120144500. Available online:
https://jphys.journaldephysique.org/en/articles/jphys/abs/1976/12/contents/contents.html
(accessed on December 1976).


\bibitem{Barber} Barber, M.N.; Guttmann, A.J.; Middlemiss, K.M.; Torrie, G.M.;  Whittington, S.G.
Some tests of scaling theory for a self-avoiding walk attached to a
surface. {\it J.Phys.A} {\bf 1978}, {\it 11}, 1833,
DOI:10.1088/0305-4470/11/9/017. Available online:
http://iopscience.iop.org/article/10.1088/0305-4470/11/9/017/meta
(accessed on September 1978).


\bibitem{DD81} Diehl, H.W.; Dietrich, S. Field-theoretical approach to static critical phenomena in semi-infinite systems.
 {\it Z.Phys.B} {\bf 1981}, {\it 42},
65-86, DOI:10.1007/2FBF01298293. Available online:
https://link.springer.com/article/10.1007\%2FBF01298293?LI=true
(accessed on March 1981).


\bibitem{BJ05} Bachmann, M.; Janke, W. Conformational Transitions of Nongrafted Polymers near an Absorbing Substrate.
{\it Phys.Rev.Lett.} {\bf 2005}, {\it 95}, 058102,
DOI:10.1103/PhysRevLett.95.058102. Available online:
https://link.aps.org/doi/10.1103/PhysRevLett.95.058102 (accessed on
29 July 2005).

\bibitem{MBJ09} M\"oddel, M.; Bachmann, M.; Janke, W. Conformational Mechanics of Polymer Adsorption Transitions at Attractive
Substrates. {\it J. Phys. Chem. B} {\bf 2009}, {\it 113}, 3314-323,
DOI:10.1021/jp808124v. Available online:
http://pubs.acs.org/doi/abs/10.1021/jp808124v (accessed on 20
February 2009).

\bibitem{MJB10} M\"oddel, M.; Janke, W.; Bachmann, M. Systematic microcanonical analyses of
polymeradsorption transitions. {\it Phys. Chem. Chem. Phys.} {\bf
2010}, {\it 12}, 11548-11554, DOI:10.1039/C002862B. Available
online:
http://pubs.rsc.org/en/content/articlelanding/2010/cp/c002862b\#!divAbstract
(accessed on 06 Aug 2010).


\bibitem{UShH01} Usatenko, Z. E.; Shpot, M. A.; Hu, Chin-Kun. Surface critical behaviour of random
systems. Ordinary transition. {\it Phys. Rev. E} {\bf 2001}, {\it
63}, 056102, DOI:10.1103/PhysRevE.63.056102. Available online:
https://link.aps.org/doi/10.1103/PhysRevE.63.056102 (accessed on 11
April 2001).

\bibitem{UH02} Usatenko, Z.; Hu, Chin-Kun. Critical behavior of semi-infinite random
systems at the special surface transition. {\it Phys. Rev. E} {\bf
2002}, {\it 65}, 066103, DOI:10.1103/PhysRevE.65.066103. Available
online: https://link.aps.org/doi/10.1103/PhysRevE.65.066103
(accessed on 13 June 2002).

\bibitem{F53} Flory, P.J. {\it Principles of Polymer Chemistry}; CornelI University Press: Ithaca, New
York, 1953; ISBN 978-0-8014-0134-3.

\bibitem{OSO97} Ohshima, Y.N.; Sakagami, H.; Okumoto, K.;
Tokoyoda, A.; Igarashi, T.; Shintaku, K.B.; Toride, S.; Sekino, H.;
Kabuto, K.; Nishio, I. Direct Measurement of Infinitesimal Depletion
Force in a Colloid-Polymer Mixture by Laser Radiation Pressure. {\it
Phys.Rev.Lett.} {\bf 1997}, {\it 78}, 3963-3966,
DOI:10.1103/PhysRevLett.78.3963. Available online:
https://journals.aps.org/prl/abstract/10.1103/PhysRevLett.78.3963
(accessed on 19 May 1997).

\bibitem{VCLY98} Verma, R.; Crocker, J.C.; Lubensky, T.C.;
Yodh, A.G. Entropic Colloidal Interactions in Concentrated DNA
Solutions. {\it Phys.Rev. Lett.} {\bf 1998}, {\it 81}, 4004-4007,
DOI:10.1103/PhysRevLett.81.4004. Available online:
https://journals.aps.org/prl/abstract/10.1103/PhysRevLett.81.4004
(accessed on 2 November 1998).


\end{thebibliography}
\end{document}